\documentclass[prd,twocolumn,showpacs,superscriptaddress,groupedaddress,nofootinbib]{revtex4-1}

\usepackage{graphicx}
\usepackage{amsmath,amssymb} 
\usepackage{color}

\usepackage{xspace}

\usepackage{tikz}
\usetikzlibrary{fit,positioning}

\def\bbbone{{\mathchoice {\rm 1\mskip-4mu l} {\rm 1\mskip-4mu l} {\rm 1\mskip-4.5mu l} {\rm 1\mskip-5mu l}}}

\newcommand{\PlotPath}{plots/}

\begin{document}

%===================================================================================================

\title{Beyond single-threshold searches:  the Event Stacking Test}
\author{Ryan Lynch}\email{ryan.lynch@ligo.org}
\author{Salvatore Vitale}
\author{Erik Katsavounidis}
\affiliation{LIGO Laboratory, Massachusetts Institute of Technology, 185 Albany St, 02139 Cambridge USA}

%===================================================================================================

\begin{abstract}

We present a new statistical test that examines the consistency of the tails of two empirical distributions at multiple thresholds.
Such distributions are often encountered in counting experiments, in physics and elsewhere, where the significance of populations of events is evaluated.
This multi-threshold approach has the effect of ``stacking'' multiple events into the tail bin of the distribution, and thus we call it the Event Stacking Test.
This test has the ability to confidently detect inconsistencies composed of multiple events, even if these events are low-significance outliers in isolation.
We derive the Event Stacking Test from first principles and show that the p-value it reports is a well-calibrated representation of noise fluctuations.
When applying this test to the detection of gravitational-wave transients in LIGO-Virgo data, we find that it performs better than or comparably to other statistical tests historically used within the gravitational-wave community.
This test is particularly well-suited for detecting classes of gravitational-wave transients that are minimally-modeled, i.e., gravitational-wave bursts.
We show that the Event Stacking Test allows us to set upper limits on the astrophysical rate-density of gravitational-wave bursts that are stricter than those set using other statistical tests by factors of up to 2 - 3.

\end{abstract}
\maketitle

%===================================================================================================
\section{Introduction}\label{Sec.Intro}

With the recent detection of gravitational-wave (GW) events such as GW150914~\cite{GW150914} and GW170817~\cite{GW170817}, we have entered an era where we expect the detection of GW transients with Advanced LIGO~\cite{StandardAdvancedLIGO} and Virgo~\cite{StandardAdvancedVirgo} to occur on a regular basis.
To date, all of the detected GW transients have been emitted by compact binary coalescence (CBC) sources that are suitably well-modeled to be detected by templated searches~\cite{PyCBCAlgo,PyCBCCode,GSTLAL}.
These detections have enabled rich scientific investigations and breakthroughs.
For example, the detection of binary black hole mergers have been used to test Einstein's theory of relativity in the strong-field regime~\cite{GW150914TestGR}.
The joint detection of the binary neutron star merger GW170817 with electromagnetic counterparts~\cite{EMFollow170817} has led to the association of short gamma-ray bursts with binary neutron star mergers~\cite{GRB170817}, has provided evidence of heavy-element nucleosynthesis~\cite{Swope,DECAM,DLT40,LasCumbres,VISTA,MASTER}, and has enabled a new procedure for measuring the Hubble parameter~\cite{HubbleConstant}.
We expect similar breakthroughs to occur when LIGO-Virgo detects non-CBC sources of GW transients, which we will refer to as GW bursts.
Examples of potential GW burst sources include the core-collapse supernovae of massive stars~\cite{Fryer:2011zz,AdvancedSNe}, neutron stars collapsing to form black holes~\cite{BurstPostMergerRemnant}, neutron star glitches~\cite{SGRR,Magnetar}, cosmic string cusps~\cite{CosmicStringTheory}, and the unknown.
While the waveforms of some of these GW burst signals can be at least partially modeled~\cite{Ott:2009,Morozova:2018glm,Dimmelmeier:2008,Yakunin:2017tus,Scheidegger:2010}, the LIGO-Virgo Collaboration has performed multi-algorithm~\cite{oLIBMethods,cwb2g,BayesWave} searches for GW bursts that are only minimally modeled to ensure that generic transient signals are reliably detected~\cite{S6Burst,Abbott:2015vir,O1AllSky,O1LongDuration}.  

Historically, the GW detection problem has been viewed as a form of outlier/anomaly detection.
The signals of GW events are superimposed onto detector noise, meaning the measured event rate in a GW analysis takes the form $\lambda_\text{total} = \lambda_\text{noise} + \lambda_\text{GW}$.
GW searches measure these event rates as a function of a search statistic $\Lambda$.
The cumulative rate of events whose measured search statistic exceeds a value $\Lambda$, $\lambda\left(\Lambda\right)$, is then given by
\begin{equation}\label{Eq.RateSuperposition}
  \lambda_\text{total}\left(\Lambda\right) = \lambda_\text{noise}\left(\Lambda\right) + \lambda_\text{GW}\left(\Lambda\right) \ \ .
\end{equation}
Typically $\lambda_\text{noise}\left(\Lambda\right) \gg \lambda_\text{GW}\left(\Lambda\right)$ for small values of $\Lambda$, meaning noise events dominate GW events in number.
However, these search statistics are constructed specifically so that the GW event rate should dominate at large values of $\Lambda$.
Thus, most detections are expected to occur in the high-$\Lambda$ tail, and we refer to events with large values of $\Lambda$ as ``loud''.
Searches for GW transients have historically evaluated the significance of events in isolation, meaning the significance of each candidate is the Poisson probability of detector noise producing at least one event exceeding the candidate's measured $\Lambda$ (e.g., see~\cite{O1BBH}).
We will refer to this process as the Loudest Event Test (LET)~\cite{LoudestEventRate}, and we will describe it in further detail in Section~\ref{Sec.Comparison}.
However, by only considering events that can be detected in isolation, we are ignoring populations of low-significance GW events that may also be located in the data.

For compact-binary coalescence (CBC) events, the astrophysical distribution of sources is believed to be well-enough understood that the GW event rate can be predicted as a function of $\Lambda$.
With such a model, we can compute the probability that any event, including low-significance ones, are part of an astrophysical population of GW events~\cite{Farr,GW150914RateSupp}.
These calculations have been performed by the LIGO-Virgo Collaboration in their first (O1) observing run in the Advanced Detector Era~\cite{O1BBH,GW150914Rate}.
However, it is difficult to do something similar for GW burst sources that are inherently unmodeled.
The expected distributions of these sources can range from point-like distributions emitted from matter-dense regions in the Milky Way Galaxy to uniform-in-volume distributions covering most of the observable universe.
Since the generic GW-burst signal morphologies and source distributions are usually assumed to be only minimally modeled, it makes sense to work in terms of null hypothesis tests where only the noise distributions must be well-modeled.
The detection statement of a null hypothesis test is that an analysis measurement is inconsistent with the measured background distribution over some region of the $\Lambda$ parameter space.

In this paper, we derive a null-hypothesis test that we will refer to as the Event Stacking Test (EST).
This test is designed to evaluate the statistical consistency of the high-$\Lambda$ tails of measured analysis and background distributions at $k$ different thresholds.
The EST evaluates the joint significance of the $k$-loudest events in the analysis data set by first ``stacking'' them into tail bins of different widths, and then comparing the statistical properties of these bins to those of the background data set.
It reports the probability that the same underlying distribution produced both the analysis and background high-$\Lambda$ tails.
As a result, the EST is able to detect up to $k$ GW events of a population, even if none of them is individually significant enough to be detected on its own.

The EST is the Poissonian analogy to the Binomial Test presented in~\cite{BinomialTest1}.
Both the EST formulation and the Binomial Test evaluate the similarity of the distributional shapes of the analysis and background events. 
However, the EST also takes into account the relative event rates of the analysis and background measurements, while the Binomial Test normalizes out these event rates.
Because GW events increase the event rate of the analysis measurement as compared to the background measurement (see Eq.~\ref{Eq.RateSuperposition}), the EST utilizes more information relevant to the GW detection problem than the Binomial Test.
We also note the p-value of the EST is calculated analytically, while the p-value of the Binomial Test historically has been computed via Monte Carlo~\cite{BinomialTest1,BinomialTest2}.
The EST is also similar in nature to the tail-targeted test presented in~\cite{CannonHannaTest} that examines the joint statistical consistency of all analysis events exceeding a certain threshold value of $\Lambda$.
We prefer the EST formulation since it always evaluates a known number of events ($k$) in an unknown region of $\Lambda$, as opposed to an unknown number of events in a known region of $\Lambda$.
For similar configurations where both tests evaluate $k$ events on average, the EST has the advantage of detecting up to $k$-event inconsistencies {\it above or below} the $\Lambda$ threshold of the referenced test, while the referenced test can detect an arbitrary-event-number inconsistency but only above the $\Lambda$ threshold.
Thus, the EST is more likely to detect low-$\Lambda$ GW events that would be missed in isolation.
Additionally, as long as all confidently-detected GW events are removed from the analysis, the EST can be repeated multiple times until statistical consistency is achieved, allowing it to detect more than $k$ GW events in practice.

In Section~\ref{Sec.Formalism} we develop the formalism needed to justify the EST and then derive it from first principles.
Then in Section~\ref{Sec.Comparison}, we compare the EST with several other standard statistical consistency tests, showing that its reported p-value is well-calibrated and that it is particularly powerful in detecting GW burst events.
We give a specific demonstration of this detection power in Section~\ref{Sec.UpperLimits}, where we show that the EST allows us to set upper limits on the astrophysical rate-density of GW burst sources that are stricter than those set using the LET.
Finally, we summarize our conclusions in Section~\ref{Sec.Conclusions}.

%===================================================================================================
\section{Formalism}\label{Sec.Formalism}

We will derive the EST in the context of the GW-transient event detection problem.
Within this context, detection is based on the analysis of timeseries data that encodes the GW strain measured by multiple interferometers, such as LIGO~\cite{StandardAdvancedLIGO}, Virgo~\cite{StandardAdvancedVirgo}, and others. 
A GW signal must be consistent with a single (astrophysical) source across all detectors.
Several consistency criteria are invoked in order to identify transient occurences of these signals within the multi-detector data streams~\cite{AllenChiSq,PyCBCBack,GSTLALLRT,oLIBMethods,cwb2g,BWcomplexity}.

The background of GW searches is commonly measured by artificially shifting the timeseries data of each detector by relative time lags so that GW events within the data are no longer found coincidently in all of the detectors.
The resulting data set is assumed to contain only noise~\cite{timeslides}.
Because the analysis data set does not undergo any time lags, it is commonly referred to as the 0-lag.
The foundation of our data model is that the occurrences of noise events in the 0-lag and background measurements are described by the same underlying Poisson process.  
Specifically, for any region of the $\Lambda$ parameter space, the number of events observed ($N$) over a duration of time ($T$) will be distributed as a Poisson distribution
  \begin{equation}\label{Eq.PoissonBack}
  P(N|\lambda,T) = \frac{1}{N!}(\lambda T)^{N} e^{-\lambda T}
  \end{equation}
where $\lambda$ is the mean event rate in this region.

\begin{figure}[t!]
    \scalebox{.7}{\includegraphics[width=0.5\textwidth]{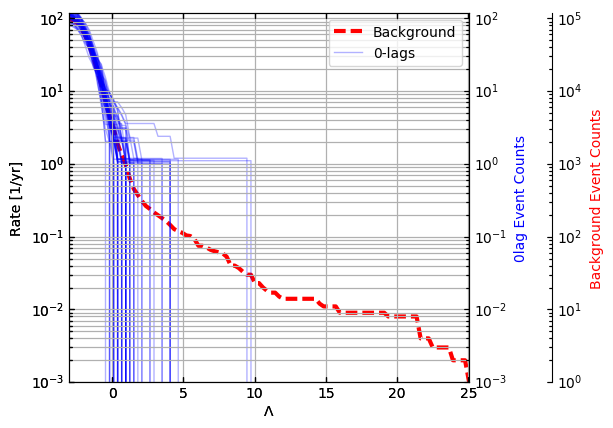}}
    \scalebox{.7}{\includegraphics[width=0.5\textwidth]{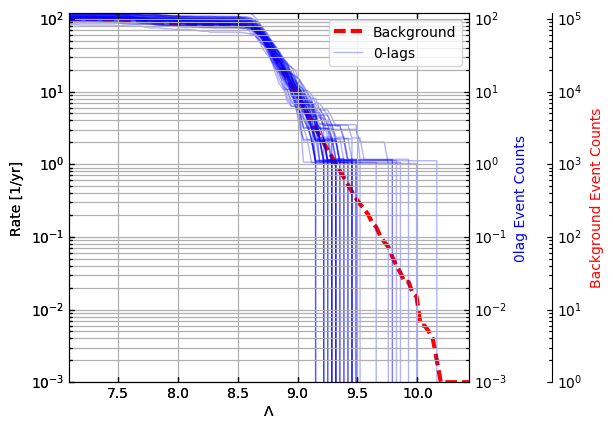}}
    \caption{Examples of 50 0-lag measurements and 1 background measurement drawn from the same noise-only event distribution for a GW burst search (top) and a CBC search (bottom).
    We sample each realization from a Poisson distribution using measurement durations $T_\text{back} = 1000$ years and $T_\text{0lag} = 1 $ year and a total event rate of 100 events per year.
    The lines trace the cumulative rate at which events exceed each value of the search statistic $\Lambda$.
    The noise-only event distributions are estimated using O1 LIGO-Virgo background measurements performed by oLIB~\cite{O1AllSky} (for GW bursts) and PyCBC~\cite{O1BBH} (for CBC).
    }
    \label{Fig.CumPlotExample}
\end{figure}

Consider a GW search that ranks events according to a search statistic $\Lambda$, where greater values of $\Lambda$ correspond to more GW-like events.  
In order to evaluate the consistency of background and 0-lag measurements made by the GW search at a value of $\Lambda$, we can compare the rate of events exceeding a given value of $\Lambda$, $\lambda\left(\Lambda\right)$, in each measurement.  
An example of this comparison is shown in Fig.~\ref{Fig.CumPlotExample} as a function of $\Lambda$.  
Doing this comparison in cumulative fashion gives a nonparametric representation of the measurement data, unlike with differential binning where the bin sizes and bin location must be specified.  
Importantly, this cumulative representation of the data has a salient feature:  the event rate must decrease monotonically as a function of $\Lambda$.  
The following sections will explore how we can use this cumulative representation of the data to quantify the level of consistency between the 0-lag and background measurements.

%========================
\subsection{Single-Threshold Significance}\label{SubSec.SingleThreshold}

\begin{figure}[t!]
  \begin{tikzpicture}
  \tikzstyle{main}=[circle, minimum size = 10mm, thick, draw =black!80, node distance = 16mm]
  \tikzstyle{connect}=[-latex, thick]
  \matrix[row sep=0.5cm,column sep=0.25cm] {
    
    \node (T0lagL) [main] {T$_\text{0lag}$};
    & & & & &
    & & & & &
    \node (T0lagR) [main,fill = black!20] {T$_\text{0lag}$}; \\
        
    \node(N0lagL) [main] {N$_\text{0lag}$};
    & & & & &
    & & & & &
    \node (N0lagR) [main] {N$_\text{0lag}$}; \\
    
    \node (lambdaL) [main] {$\lambda$};
    & & & & &
    & & & & &
    \node (lambdaR) [main] {$\lambda$}; \\

    \node(NbackL) [main] {N$_\text{back}$};
    & & & & &
    & & & & &
    \node (NbackR) [main,fill = black!20] {N$_\text{back}$}; \\
    
    \node (TbackL) [main] {T$_\text{back}$};
    & & & & &
    & & & & &
    \node (TbackR) [main,fill = black!20] {T$_\text{back}$}; \\
  };
  \path[->]
    (lambdaL) edge[thick] (N0lagL)
    (lambdaL) edge[thick] (NbackL)
    (T0lagL) edge[thick] (N0lagL)
    (TbackL) edge[thick] (NbackL)
    
    (lambdaR) edge[thick] (N0lagR)
    (lambdaR) edge[thick] (NbackR)
    (T0lagR) edge[thick] (N0lagR)
    (TbackR) edge[thick] (NbackR)
    ;
  \end{tikzpicture}
  \caption{The DAG describing the probabilistic dependencies of variables in the single-threshold test.
  The variables are denoted by nodes, and the arrows point from cause to effect.
  Shaded nodes represent variables that have been measured and have fixed values, while the unshaded nodes represent variables that are unconstrained.
  The left DAG shows the state of knowledge before any measurements are made, while the right DAG shows the state of knowledge after the relevant measurements are made.
  The explicit probabilistic dependencies implied by each DAG can be extracted using the theory of D-separation~\cite{DSeparation,Bishop}.
       }
  \label{Fig.DAGs_single}
\end{figure}
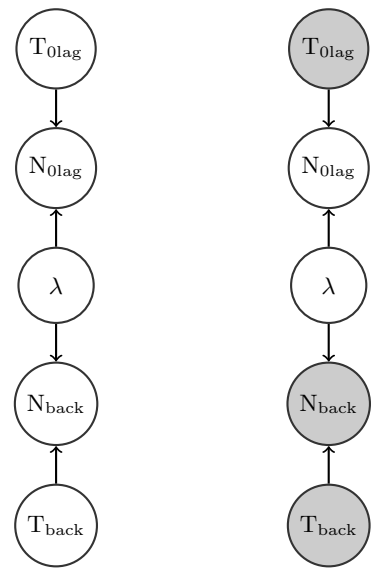

Assuming Poissonity, it is straightforward to evaluate the statistical consistency of any 0-lag and background measurements at a single threshold value of $\Lambda$.
Because the inconsistencies of interest to us are excesses of events in the 0-lag, the p-value we will calculate is the false-alarm-probability (FAP).  
The FAP at a threshold of $\Lambda$ is given by 
  \begin{equation}\label{Eq.FAP_single}
  \begin{split}
  \text{FAP}_\Lambda(N_\text{0lag}^*) &= P(N_\text{0lag} \geq N_\text{0lag}^* | \vec{\theta}) \\
					       &=\sum_{N_\text{0lag} = N_\text{0lag}^*}^\infty P(N_\text{0lag} | \vec{\theta}) \\
					       &= 1 - \sum_{N_\text{0lag} = 0}^{N_\text{0lag}^* - 1} P(N_\text{0lag} | \vec{\theta})
  \end{split}
  \end{equation}
where $N_\text{0lag}$ is the number of 0-lag events exceeding $\Lambda$, $N_\text{0lag}^*$ denotes minimum value of $N_\text{0lag}$ that will be considered a significant excess, and $\vec{\theta}$ represents the full set of conditional parameters that have been measured.  In practice, one commonly sets $N_\text{0lag}^*$ equal to the number 0-lag events measured to be exceeding $\Lambda$ and then uses Eq.~\ref{Eq.FAP_single} to calculate the significance of this 0-lag measurement.

In order to calculate Eq.~\ref{Eq.FAP_single}, we need to first specify the observed quantities $\vec{\theta}$ and then find the functional form of $P(N_\text{0lag} | \vec{\theta})$.  
We can model the probabilistic dependencies of the parameters involved in these steps using a probabilistic directed acyclic graphical model (DAG).  
While a detailed formulation of probabilistic DAGs (also commonly referred to as Bayesian Networks or Bayesian Hierarchical Models) can be found in~\cite{DSeparation,Bishop}, the basic premise is fairly intuitive:  nodes represent variables, and arrows describe the cause-and-effect relationship between pairs of variables by pointing from cause to effect.  
Unshaded nodes represent free variables, while shaded nodes represent measured variables.

The DAG for this single-threshold problem is shown in Fig.~\ref{Fig.DAGs_single}.  
The interpretation of the DAG is as follows:  there is an underlying Poisson process with noise event rate $\lambda \equiv \lambda\left(\Lambda\right)$ that, along with 0-lag ($T_\text{0lag}$) and background ($T_\text{back}$) measurement durations, generates a number of 0-lag ($N_\text{0lag}$) and background ($N_\text{back}$) events exceeding the threshold $\Lambda$.  
Let us then consider the scenario where we have measured $T_\text{0lag}$, $T_\text{back}$, and $N_\text{back}$.
In order to solve Eq.~\ref{Eq.FAP_single}, we need to find the functional form of
  \begin{equation}\label{Eq.N0lagSingle}
  \begin{split}
  P(N_\text{0lag} | \vec{\theta}) &= \ P(N_\text{0lag}|T_\text{0lag},T_\text{back},N_\text{back}) \\
				  &= \int_0^\infty d\lambda \ P(N_\text{0lag},\lambda|T_\text{0lag},T_\text{back},N_\text{back}) \ \ . 
  \end{split}
  \end{equation}
To accomplish this, we can use the theory of D-separation~\cite{DSeparation,Bishop} on the DAG to find the following conditional independencies: (1) $N_\text{0lag} \perp N_\text{back}, T_\text{back} | \lambda, T_\text{0lag}$, and (2) $\lambda \perp T_\text{0lag} | N_\text{back},T_\text{back}$.
These conditional independencies allow us to write
%  \begin{widetext}
  \begin{equation}\label{Eq.FactorSingleThresh}
  \begin{split}
  P(N_\text{0lag},\lambda &| T_\text{0lag},T_\text{back},N_\text{back}) \\
			  &= \ P(N_\text{0lag}|\lambda, T_\text{0lag},T_\text{back},N_\text{back}) \\
			  & \ \ \ \ \ \ \ \cdot P(\lambda|T_\text{0lag},T_\text{back},N_\text{back}) \\
			  &= \ P(N_\text{0lag}|\lambda, T_\text{0lag}) \ P(\lambda|N_\text{back}, T_\text{back}) \ \ .
  \end{split}
  \end{equation}
%  \end{widetext}

One approach to solving Eq.~\ref{Eq.N0lagSingle} is to estimate $\lambda$ using the maximum-likelihood estimator $\hat{\lambda}_\text{ML} = \frac{N_\text{back}}{T_\text{back}}$.
With this choice, we can write $P(\lambda|N_\text{back},T_\text{back}) = \delta(\lambda-\hat{\lambda}_\text{ML})$.
Substituting this delta function into Eq.~\ref{Eq.FactorSingleThresh} and integrating over $\lambda$, we find
  \begin{equation}
  \begin{split}
  P_\text{ML}(N_\text{0lag} | T_\text{0lag},T_\text{back},N_\text{back}) = \ P(N_\text{0lag}|\hat{\lambda}_\text{ML},T_\text{0lag}) \ \ . 
  \end{split}
  \end{equation}
The right-hand side of this expression can be identified as the Poisson distribution defined in Eq.~\ref{Eq.PoissonBack}.
We thus find the ``maximum-likelihood'' FAP estimate to be
  \begin{equation}\label{Eq.FAP_single_ML}
  \begin{split}
  \text{FAP}_{\Lambda,\text{ML}} & (N_\text{0lag}^*) \\ 
				 &= 1 - \sum_{N_\text{0lag} = 0}^{N_\text{0lag}^* - 1} P_\text{ML}(N_\text{0lag} | T_\text{0lag},T_\text{back},N_\text{back})
  \end{split}
  \end{equation}
where
  \begin{equation}\label{Eq.P_ML}
  \begin{split}
  P_\text{ML}(N_\text{0lag} &| T_\text{0lag},T_\text{back},N_\text{back}) \\ 
			    &= \frac{1}{N_\text{0lag}!}(\hat{\lambda}_\text{ML} T_\text{0lag})^{N_\text{0lag}} e^{-\hat{\lambda}_\text{ML} T_\text{0lag}} \ \ . 
  \end{split}
  \end{equation}
  
Instead of using a maximum-likelihood estimator for $\lambda$, we can instead choose to adopt the Bayesian approach of marginalizing over the unobserved $\lambda$.
We can apply Bayes' theorem to Eq.~\ref{Eq.FactorSingleThresh} to obtain
  \begin{widetext}
  \begin{equation}
  P(N_\text{0lag},\lambda|T_\text{0lag},T_\text{back},N_\text{back}) = \ \frac{P(N_\text{0lag}|\lambda, T_\text{0lag}) \ P(N_\text{back}|\lambda, T_\text{back}) \ P(\lambda)}{\int_0^\infty d\lambda \ P(N_\text{back}|\lambda, T_\text{back}) \ P(\lambda)}
  \end{equation}
  \end{widetext}
where we have again used D-separation on the DAG to find the independency $ \lambda \perp T_\text{back}$.
Substituting this expression into Eq.~\ref{Eq.N0lagSingle}, we find that we need to solve two integrals:
  \begin{widetext}
  \begin{equation}
  P(N_\text{0lag}|T_\text{0lag},T_\text{back},N_\text{back}) = \frac{\int_0^\infty d\lambda \ P(N_\text{0lag}|\lambda, T_\text{0lag}) \ P(N_\text{back}|\lambda, T_\text{back}) \ P(\lambda)}{\int_0^\infty d\lambda \ P(N_\text{back}|\lambda, T_\text{back}) \ P(\lambda)} \ \ . 
  \end{equation}
  \end{widetext}
Both integrands involve the product of Poisson distributions and priors on $\lambda$, and both integrals can be solved analytically by choosing $P(\lambda)$ to be a Gamma distribution.
We will solve this integration explicitly for two such ``uninformative'' priors:  a uniform prior where $P_\text{uni}(\lambda) \propto \lambda^0$, and a Jeffreys prior where $P_\text{Jef}(\lambda) \propto \lambda^{-\frac{1}{2}}$.
For these priors, we find
  \begin{widetext}
  \begin{equation}\label{Eq.P_uniform}
  P_\text{uni}(N_\text{0lag}|T_\text{0lag},T_\text{back},N_\text{back}) = \frac{(N_{\text{back}}+N_{\text{0lag}})!}{N_{\text{back}}! ~ N_{\text{0lag}}!} \cdot \frac{T_{\text{0lag}}^{N_{\text{0lag}}} ~ T_{\text{back}}^{N_{\text{back}}+1}}{(T_{\text{back}} + T_{\text{0lag}})^{N_{\text{back}} + N_{\text{0lag}} + 1}} 
  \end{equation}
  \begin{equation}\label{Eq.P_Jeffreys}
  P_\text{Jef}(N_\text{0lag}|T_\text{0lag},T_\text{back},N_\text{back}) = \frac{(N_{\text{back}}+N_{\text{0lag}} - \frac{1}{2})!}{(N_{\text{back}}-\frac{1}{2})! ~ N_{\text{0lag}}!} \cdot \frac{T_{\text{0lag}}^{N_{\text{0lag}}} ~ T_{\text{back}}^{N_{\text{back}}+\frac{1}{2}}}{(T_{\text{back}} + T_{\text{0lag}})^{N_{\text{back}} + N_{\text{0lag}} + \frac{1}{2}}} . 
  \end{equation}
  \end{widetext}
Using these expressions, we find the ``uniform FAP'' and the ``Jeffreys FAP'' estimates to be
\\
  \begin{equation}\label{Eq.FAP_single_uniform}
  \begin{split}
  \text{FAP}_{\Lambda,\text{uni}}&(N_\text{0lag}^*) = \\
				 &1 - \sum_{N_\text{0lag} = 0}^{N_\text{0lag}^* - 1} P_\text{uni}(N_\text{0lag}|N_\text{back},T_\text{back},T_\text{0lag}) \\
  \end{split}
  \end{equation}
and
  \begin{equation}\label{Eq.FAP_single_Jeffreys}
  \begin{split}
  \text{FAP}_{\Lambda,\text{Jef}}&(N_\text{0lag}^*) = \\
				 &1 - \sum_{N_\text{0lag} = 0}^{N_\text{0lag}^* - 1} P_\text{Jef}(N_\text{0lag}|N_\text{back},T_\text{back},T_\text{0lag}) \ \ ,
  \end{split}
  \end{equation}
respectively.

%========================
\subsection{Multi-Threshold Significance}\label{SubSec.MultipleThresholds}

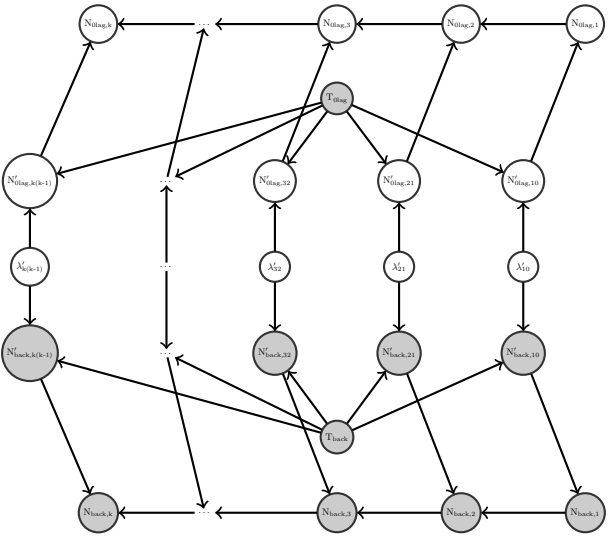
\begin{figure}[t!]
  \begin{tikzpicture}[thick,scale=0.6, every node/.style={scale=0.4}]
  \tikzstyle{main}=[circle, minimum size = 10mm, thick, draw =black!80, node distance = 16mm]
  \tikzstyle{connect}=[-latex, thick]
  \matrix[row sep=0.5cm,column sep=0.25cm] {
    
    &
    \node (N0lagN) [main] {N$_\text{0lag,k}$};
    & & &
    \node (N0lagdots) {$\dots$};
    & & &
    \node (N0lag3) [main] {N$_\text{0lag,3}$};
    & &
    \node (N0lag2) [main] {N$_\text{0lag,2}$};
    & &
    \node (N0lag1) [main] {N$_\text{0lag,1}$}; \\

    & & & & & & &
    \node (T0lag) [main,fill = black!20] {T$_\text{0lag}$}; \\
    
    \node(N0lagNprime) [main] {N$_\text{0lag,k(k-1)}'$};
    & & &
    \node(N0lagdotsprime) {$\dots$};
    & & &
    \node (N0lag3prime) [main] {N$_\text{0lag,32}'$};
    & &
    \node (N0lag2prime) [main] {N$_\text{0lag,21}'$};
    & &
    \node (N0lag1prime) [main] {N$_\text{0lag,10}'$}; \\
    
    \node (lambdaNprime) [main] {$\lambda_\text{k(k-1)}'$};
    & & &
    \node (lambdadotsprime) {$\dots$};
    & & &
    \node (lambda3prime) [main] {$\lambda_\text{32}'$};
    & &
    \node (lambda2prime) [main] {$\lambda_\text{21}'$};
    & &
    \node (lambda1prime) [main] {$\lambda_\text{10}'$}; \\

    \node(NbackNprime) [main,fill = black!20] {N$_\text{back,k(k-1)}'$};
    & & &
    \node(Nbackdotsprime) {$\dots$};
    & & &
    \node (Nback3prime) [main,fill = black!20] {N$_\text{back,32}'$};
    & &
    \node (Nback2prime) [main,fill = black!20] {N$_\text{back,21}'$};
    & &
    \node (Nback1prime) [main,fill = black!20] {N$_\text{back,10}'$}; \\
    
    & & & & & & &
    \node (Tback) [main,fill = black!20] {T$_\text{back}$}; \\
    
    &
    \node (NbackN) [main,fill = black!20] {N$_\text{back,k}$};
    & & &
    \node (Nbackdots) {$\dots$};
    & & &
    \node (Nback3) [main,fill = black!20] {N$_\text{back,3}$};
    & &
    \node (Nback2) [main,fill = black!20] {N$_\text{back,2}$};
    & &
    \node (Nback1) [main,fill = black!20] {N$_\text{back,1}$}; \\
  };
  \path[->]
    (lambdaNprime) edge[thick] (N0lagNprime)
    (N0lagNprime) edge[thick] (N0lagN)
    
    (lambdadotsprime) edge[thick] (N0lagdotsprime)
    (N0lagdotsprime) edge[thick] (N0lagdots)
    (N0lagdots) edge[thick] (N0lagN)
    
    (lambda3prime) edge[thick] (N0lag3prime)
    (N0lag3prime) edge[thick] (N0lag3)
    (N0lag3) edge[thick] (N0lagdots)
    
    (lambda2prime) edge[thick] (N0lag2prime)
    (N0lag2prime) edge[thick] (N0lag2)
    (N0lag2) edge[thick] (N0lag3)
    
    (lambda1prime) edge[thick] (N0lag1prime)
    (N0lag1prime) edge[thick] (N0lag1)
    (N0lag1) edge[thick] (N0lag2)
    
    (lambdaNprime) edge[thick] (NbackNprime)
    (NbackNprime) edge[thick] (NbackN)
    
    (lambdadotsprime) edge[thick] (Nbackdotsprime)
    (Nbackdotsprime) edge[thick] (Nbackdots)
    (Nbackdots) edge[thick] (NbackN)
    
    (lambda3prime) edge[thick] (Nback3prime)
    (Nback3prime) edge[thick] (Nback3)
    (Nback3) edge[thick] (Nbackdots)
    
    (lambda2prime) edge[thick] (Nback2prime)
    (Nback2prime) edge[thick] (Nback2)
    (Nback2) edge[thick] (Nback3)
    
    (lambda1prime) edge[thick] (Nback1prime)
    (Nback1prime) edge[thick] (Nback1)
    (Nback1) edge[thick] (Nback2)
    
    (T0lag) edge[thick] (N0lagNprime)
    %(T0lag) edge[thick] (N0lagN)
    (T0lag) edge[thick] (N0lagdotsprime)
    %(T0lag) edge[thick] (N0lagdots)
    (T0lag) edge[thick] (N0lag3prime)
    %(T0lag) edge[thick] (N0lag3)
    (T0lag) edge[thick] (N0lag2prime)
    %(T0lag) edge[thick] (N0lag2)
    (T0lag) edge[thick] (N0lag1prime)
    %(T0lag) edge[thick] (N0lag1)
    
    (Tback) edge[thick] (NbackNprime)
    %(Tback) edge[thick] (NbackN)
    (Tback) edge[thick] (Nbackdotsprime)
    %(Tback) edge[thick] (Nbackdots)
    (Tback) edge[thick] (Nback3prime)
    %(Tback) edge[thick] (Nback3)
    (Tback) edge[thick] (Nback2prime)
    %(Tback) edge[thick] (Nback2)
    (Tback) edge[thick] (Nback1prime)
    %(Tback) edge[thick] (Nback1)
    ;
  \end{tikzpicture}
  \caption{The DAG describing the probabilistic dependencies of variables in the k-threshold test.
  The variables are denoted by nodes, and the arrows point from cause to effect.
  Shaded nodes represent variables that have been measured and have fixed values, while the unshaded nodes represent variables that are unconstrained.
  Primed variables represent the quantities as measured between the thresholds, while un-primed variable represent the quantities as measured cumulatively above the threshold. 
  The explicit probabilistic dependencies implied by each DAG can be extracted using the theory of D-separation~\cite{DSeparation,Bishop}.
       }
  \label{Fig.DAG_multiple_N_observed}
\end{figure}

If we choose to make measurements a multiple thresholds, $\vec{\Lambda}$, instead of just a single threshold, the measurement at each threshold is capable of reporting statistical inconsistencies. 
As a result, multiple-threshold consistency tests are biased towards reporting inconsistencies between 0-lag and background measurements more frequently than single-threshold consistency tests.
For this reason, we must take care to properly calibrate the FAPs that are reported when making multiple measurements.
Similarly to Sec~\ref{SubSec.SingleThreshold}, let $\vec{N}_\text{0lag}^*$ denote the minimum number of 0-lag events exceeding the thresholds $\vec{\Lambda}$ that will be considered a significant excess.
Instead of Eq.~\ref{Eq.FAP_single}, the FAP of observing a significant excess at any of the thresholds will be
  \begin{equation}\label{Eq.FAP_multiple_full}
  \begin{split}
  \text{FAP}_{\vec{\Lambda}}(\vec{N}_\text{0lag}^*) &= P(\textbf{any} \ N_\text{0lag,i}  \geq N_\text{0lag,i}^* \in \vec{N}_\text{0lag}^* | \vec{\theta}) \\
								   &= 1 - P(\textbf{all} \ N_\text{0lag,i}  < N_\text{0lag,i}^* \in \vec{N}_\text{0lag}^* | \vec{\theta}) \\
  \end{split}
  \end{equation}
where $N_\text{0lag,i} \in \vec{N}_\text{0lag}$ represents the number of 0-lag events exceeding the threshold $\Lambda_i \in \vec{\Lambda}$ and $\vec{\theta}$ again represents the full set of conditional parameters that have been measured.

Let us consider the problem of calculating the statistical consistency at $k$ value-ordered thresholds $\vec{\Lambda}$ in the cumulative representation.
We choose $\vec{N}_\text{0lag}^*$ so that 
  \begin{equation}\label{Eq.CriticalThresholds}
  \text{FAP}_{\Lambda_i}(N_\text{0lag,i}^*) \leq \text{FAP}_\text{single} < \text{FAP}_{\Lambda_i}(N_\text{0lag,i}^*-1)
  \end{equation}
for all thresholds $\Lambda_i \in \vec{\Lambda}$.
This means that each $N_\text{0lag,i}^*$ is chosen so that the FAP at threshold $\Lambda_i$ (calculated using Eq.~\ref{Eq.FAP_single_ML},~\ref{Eq.FAP_single_uniform}, or~\ref{Eq.FAP_single_Jeffreys} depending on the choice of prior) is closest to some FAP$_\text{single}$ without exceeding it.
As a result, we can view FAP$_\text{single}$ as our single-threshold reference FAP that every threshold in $\vec{\Lambda}$ tests for.
Under this point of view, we write
  \begin{equation}\label{Eq.FAP_multiple_ETF}
  \text{FAP}_{\vec{\Lambda}}(\vec{N}_\text{0lag}^*) = \text{ETF}(\vec{N}_\text{0lag}^*)\cdot\text{FAP}_\text{single} \ \ .
  \end{equation}
for some $\text{ETF} \geq 0$.
We refer to the variable ETF as the ``effective trials factor'' because it is the coefficient linking a single-threshold FAP to a multiple-threshold FAP, with each threshold representing a ``trial'' that could produce an excess of the desired significance.
ETF tells us the actual probability of finding an inconsistency of significance FAP$_\text{single}$, accounting for the multiple measurements we have performed by using multiple thresholds. 
In the case of independent measurements, the Bonferroni correction~\cite{Bonferroni} tells us that we should set ETF equal to the number of thresholds.
Nevertheless, the word ``effective'' denotes that ETF will not correspond to the exact number of thresholds we have used if there is any non-zero dependence between the measurements at each threshold.

In the cumulative representation of the measurements depicted in Fig.~\ref{Fig.CumPlotExample}, the measurements are not independent as a result of the constraint that the cumulative event rate $\lambda\left(\Lambda\right)$ must decrease monotonically as a function of $\Lambda$.
In order to calculate $\text{FAP}_{\vec{\Lambda}}$, we need to understand how this monotonicity constraint affects the relationship among the $k$ thresholds $\vec{\Lambda}$.
Let us order the thresholds from largest to smallest so that $\Lambda_i \geq \Lambda_{i+1} \ \forall \ \Lambda_i,\Lambda_{i+1} \in \vec{\Lambda}$.
The monotonicity of the cumulative representation tells us that any event, 0-lag or background, that exceeds the threshold $\Lambda_i$ must also exceed all thresholds $\Lambda_j$ for $j>i$.
Thus, the elements of the event-number vector $\vec{N}$ at thresholds $\vec{\Lambda}$ are correlated with each other for both 0-lag and background measurements.
We can explain this dependence quantitatively by introducing a new set of variables:  if $N_i$ and $\lambda_i \equiv \lambda\left(\Lambda_i\right)$ are the number of events and rate of events exceeding threshold $\Lambda_i$, then we define $N'_{i(i-1)}$ and $\lambda'_{i(i-1)}$ to be the number of events and the rate of events exceeding threshold $\Lambda_i$ but not threshold $\Lambda_{i-1}$.
Then, the event rate $\lambda'_{i(i-1)}$ along with the measurement duration $T$ determines $N'_{i(i-1)}$ according to a Poisson distribution.
The number of events we measure exceeding threshold $\Lambda_i$ is given by $N_i = N_{i-1} + N'_{i(i-1)}$.
These cause-and-effect relationships allow us to depict the dependence among all variables with a probabilistic DAG, which we illustrate in Fig.~\ref{Fig.DAG_multiple_N_observed}.

We can solve for FAP$_{\vec{\Lambda}}$ by rewriting Eq.~\ref{Eq.FAP_multiple_full} as
  \begin{equation}\label{Eq.FAP_multiple_sum}
  \begin{split}
  \text{FAP}&_{\vec{\Lambda}}(\vec{N}_\text{0lag}^*) \\
	    &= 1 - \sum_{\vec{N}_\text{0lag} < \vec{N}_\text{0lag}^*} P(\vec{N}_\text{0lag} | \vec{N}_\text{back},\vec{N}_\text{back}',T_\text{back},T_\text{0lag})
  \end{split}
  \end{equation}
where we note that our calculation is conditioned on all measurements available to us: the number of background events exceeding the thresholds and the durations of both the background and 0-lag measurements.
We can expand the summed probability as:
  \begin{widetext}
  \begin{equation}\label{Eq.P_of_vecN_long_1}
  \begin{split}
  P(\vec{N}_\text{0lag} | \vec{N}_\text{back}, \vec{N}_\text{back}',T_\text{back},T_\text{0lag}) &= \sum_{\vec{N}_\text{0lag}'} \int_0^\infty d\vec{\lambda} \ P(\vec{N}_\text{0lag},\vec{N}_\text{0lag}',\vec{\lambda} | \vec{N}_\text{back},\vec{N}_\text{back}',T_\text{back},T_\text{0lag}) \\
    &= \sum_{\vec{N}_\text{0lag}'} \int_0^\infty d\vec{\lambda} \ P(\vec{N}_\text{0lag} | \vec{N}_\text{0lag}',\vec{\lambda}, \vec{N}_\text{back},\vec{N}_\text{back}',T_\text{back},T_\text{0lag}) \\ 
    & \ \ \ \ \ \ \ \ \ \ \ \ \cdot P(\vec{N}_\text{0lag}'| \vec{\lambda}, \vec{N}_\text{back},\vec{N}_\text{back}',T_\text{back},T_\text{0lag}) \ P(\vec{\lambda} | \vec{N}_\text{back},\vec{N}_\text{back}',T_\text{back},T_\text{0lag}) \ \ .
  \end{split}
  \end{equation}
  \end{widetext}
Applying D-separation to the DAG illustrated in Fig.~\ref{Fig.DAG_multiple_N_observed}, we find the following conditional independencies:  (1) $\vec{N}_\text{0lag} \perp \vec{\lambda}, \vec{N}_\text{back},\vec{N}_\text{back}',T_\text{back},T_\text{0lag}  \ | \ \vec{N}_\text{0lag}'$; (2) $\vec{N}_\text{0lag}' \perp \vec{N}_\text{back},\vec{N}_\text{back}',T_\text{back} \ | \ \vec{\lambda}, T_\text{0lag}$; (3) $\vec{\lambda} \perp \vec{N}_\text{back}, T_\text{0lag} \ | \ \vec{N}_\text{back}', T_\text{back}$.
These conditional independencies allow us to simplify Eq.~\ref{Eq.P_of_vecN_long_1} to
  \begin{widetext}
  \begin{equation}\label{Eq.P_of_vecN_long_2}
  \begin{split}
  P(\vec{N}_\text{0lag} | \vec{N}_\text{back}, \vec{N}_\text{back}',T_\text{back},T_\text{0lag}) &= \sum_{\vec{N}_\text{0lag}'} \int_0^\infty d\vec{\lambda} \ P(\vec{N}_\text{0lag} | \vec{N}_\text{0lag}') \ P(\vec{N}_\text{0lag}'| \vec{\lambda}, T_\text{0lag}) \ P(\vec{\lambda} |\vec{N}_\text{back}',T_\text{back}) \\
  &= \sum_{\vec{N}_\text{0lag}'} \int_0^\infty d\vec{\lambda} \ \left[ \prod_{i=1}^k P(N_{\text{0lag},i} | \lbrace N_{\text{0lag},j} : j < i \rbrace ,\vec{N}_\text{0lag}') \right] \\
  & \ \ \ \ \ \ \ \ \ \ \ \ \cdot P(\vec{N}_\text{0lag}'| \vec{\lambda}, T_\text{0lag}) \ P(\vec{\lambda} |\vec{N}_\text{back}',T_\text{back})
  \end{split}
  \end{equation}
  \end{widetext}
where we have conditionally factored $P(\vec{N}_\text{0lag} | \vec{N}_\text{0lag}')$ in the second line.
We can once again use D-separation on the DAG of Fig.~\ref{Fig.DAG_multiple_N_observed} to find three more conditional independencies:  (4) $N_{\text{0lag},i} \perp N_{\text{0lag},j < i-1}, N_{\text{0lag},j(j-1) \neq i(i-1)}' \ | \ N_{\text{0lag},i-1},N_{\text{0lag}, i(i-1)}'$; (5) $N_{\text{0lag},i(i-1)}' \perp \lambda_{j(j-1) \neq i(i-1)}, N_{\text{0lag},j(j-1) \neq i(i-1)}'$ $| \ \lambda_{i(i-1)}, T_\text{0lag}$; (6) $\lambda_{i(i-1)} \perp \lambda_{j(j-1)\neq i(i-1)},$ $ N_{\text{back},j(j-1)\neq i(i-1)}'$ $| \ N_{\text{back},i(i-1)}', T_\text{back}$.
These additional conditional independencies again let us simplify the equation of interest to
  \begin{widetext}
  \begin{equation}\label{Eq.P_of_vecN_long_3}
  \begin{split}
  P(\vec{N}_\text{0lag} | \vec{N}_\text{back}, \vec{N}_\text{back}',T_\text{back},T_\text{0lag}) &= \sum_{\vec{N}_\text{0lag}'} \ \left[ \prod_{i=1}^k P(N_{\text{0lag},i} | N_{\text{0lag},i-1}, N_{\text{0lag},i(i-1)}') \right] \\
  & \ \ \ \ \ \ \ \ \ \ \ \ \cdot \int_0^\infty d\vec{\lambda} \ \left[ \prod_{j=1}^k P(N_{\text{0lag},j(j-1)}' | \lambda_{j(j-1)}, T_\text{0lag}) \ P(\lambda_{j(j-1)} | N_{\text{back},j(j-1)}',T_\text{back}) \right]
  \end{split}
  \end{equation}
  \end{widetext}
where we note that $N_{\text{0lag},0}$ is a placeholder variable that can be set to 0.
First, we note that $P\left(N_{\text{0lag},i} | N_{\text{0lag},i-1}, N_{\text{0lag},i(i-1)}'\right) = \bbbone \left( N_{\text{0lag},i} = N_{\text{0lag},i-1} + N_{\text{0lag},i(i-1)}' \right)$.
Second, we also note that each product term in the second integral is equivalent to the right-hand side of Eq.~\ref{Eq.FactorSingleThresh}.  
Thus, using the discussion of Sec.~\ref{SubSec.SingleThreshold} to integrate over $\vec{\lambda}$, we can rewrite Eq.~\ref{Eq.P_of_vecN_long_3} as
  \begin{widetext}
  \begin{equation}\label{Eq.P_of_vecN_long_4}
  \begin{split}
  P(\vec{N}_\text{0lag} | & \vec{N}_\text{back}, \vec{N}_\text{back}',T_\text{back},T_\text{0lag}) \\
			  &= \sum_{\vec{N}_\text{0lag}'} \prod_{i=1}^k \bbbone \left( N_{\text{0lag},i} = N_{\text{0lag},i-1} + N_{\text{0lag},i(i-1)}' \right) P_\text{pri}(N_{\text{0lag},i(i-1)}' | T_\text{0lag}, T_\text{back}, N_{\text{back},i(i-1)}') \\
  \end{split}
  \end{equation}
  \end{widetext}
where $\text{pri} \in \lbrace \text{ML}, \text{Jef}, \text{uni} \rbrace$ represents our choice of prior on $\lambda$.
Finally, completing the summation, we find
  \begin{equation}\label{Eq.P_of_vecN_long_final}
  \begin{split}
  & P(\vec{N}_\text{0lag} | \vec{N}_\text{back}, \vec{N}_\text{back}',T_\text{back},T_\text{0lag}) \\
  &= \prod_{i=1}^k P_\text{pri}(N_{\text{0lag},i} - N_{\text{0lag},i-1} | T_\text{0lag}, T_\text{back}, N_{\text{back},i(i-1)}') \ \ .
  \end{split}
  \end{equation}
This expression can be used to complete the sum in Eq.~\ref{Eq.FAP_multiple_sum}, giving us our multi-threshold FAP.
The effective trials factor ETF is then trivially found using Eq.~\ref{Eq.FAP_multiple_ETF}, which amounts to dividing by FAP$_\text{single}$.

We note that we can calculate the sum in Eq.~\ref{Eq.FAP_multiple_sum} in a scalable and efficient manner using the following algorithm.
First, we sum over $N_{\text{0lag},1}$, noting that the sum will also be dependent upon $N_{\text{0lag},2}$:
  \begin{equation}
  \begin{split}
  & m_{2}(N_{\text{0lag},2}) \\
  &= \sum_{N_{\text{0lag},1}=0}^{N_{\text{0lag},1}^*-1} P_\text{pri}(N_{\text{0lag},2} - N_{\text{0lag},1} | T_\text{0lag}, T_\text{back}, N_{\text{back},21}') \\
  & \ \ \ \ \ \ \ \ \ \ \ \ \cdot P_\text{pri}(N_{\text{0lag},1} | T_\text{0lag}, T_\text{back}, N_{\text{back},10}') \ \ .
  \end{split}
  \end{equation}
We can then iterate over the indices of summation, generating the following recursive relationship for $1 < i < k$:
  \begin{widetext}
  \begin{equation}
  m_{i+1}(N_{\text{0lag},i+1}) = \sum_{N_{\text{0lag},i}=0}^{N_{\text{0lag},i}^*-1} P_\text{pri}(N_{\text{0lag},i+1} - N_{\text{0lag},i} | T_\text{0lag}, T_\text{back}, N_{\text{back},(i+1)i}') \ m_{i}(N_{\text{0lag},i}) \ \ .
  \end{equation}
  \end{widetext}
Carrying this through to the $k^{th}$ threshold, we find that Eq.~\ref{Eq.FAP_multiple_sum} can be rewritten as:
  \begin{equation}\label{Eq.FAP_multiple_message}
  \text{FAP}_{\vec{\Lambda}}(\vec{N}_\text{0lag}^*) = 1 - \sum_{N_{\text{0lag},k}=0}^{N_{\text{0lag},k}^*-1}  m_{k}(N_{\text{0lag},k}) \ \ .
  \end{equation}
Because there are $k$ sums, each depending on at most 2 of the summation variables, the computational complexity of this algorithm is only $\mathcal{O} \left( k \left[ \max{\lbrace \vec{N}_\text{0lag}^* \rbrace} \right]^2 \right)$.
Thus, the exact calculation of Eq.~\ref{Eq.FAP_multiple_message} is scalable in both the number of thresholds $k$ and the level of significance, which determines $\max{\lbrace \vec{N}_\text{0lag}^* \rbrace}$.
  
%========================
\subsection{The Event Stacking Test}\label{SubSec.EST}

We will now discuss solving Eq.~\ref{Eq.FAP_multiple_message} for a specific selection of $k$ thresholds, which will form the tail-targeted statistical consistency test that we will refer to as the Event Stacking Test (EST).
The cumulative representation of data, as depicted in Fig.~\ref{Fig.CumPlotExample}, is appropriate for checking for distributional inconsistencies in the tails of measurements since all events measured to exceed a $\Lambda$ threshold are ``stacked'' into the tail bin defined by $\Lambda$.
Choosing to make measurements at the multiple thresholds of $\vec{\Lambda}$ is equivalent to testing the distributional consistency of the high-$\Lambda$ tail for several binnings, where the value of each $\Lambda_i$ defines the size of the tail bin.
Specifically, large values of $\Lambda_i$ correspond to short tails with few events, while small values of $\Lambda_i$ allow for much wider tail regions that encompass more events.
We note that Eq.~\ref{Eq.P_of_vecN_long_final} is the product of the probabilities of observing each $N_\text{0lag,i(i-1)}' \in \vec{N}_\text{0lag}'$, which is equivalent to calculating the joint probability of a 0-lag measurement using the differentially-binned event counts $\vec{N}_\text{0lag}'$.
The transformation that maps a set of differentially-binned event counts into a multi-threshold test of the measurement's tail is performed by changing to the cumulative representation of event counts $\vec{N}_\text{0lag}$ in the summation bounds of Eq.~\ref{Eq.FAP_multiple_sum}.

%===
\subsubsection{Choosing optimal thresholds}

\begin{figure}[t!]
    \scalebox{.9}{\includegraphics[width=0.5\textwidth]{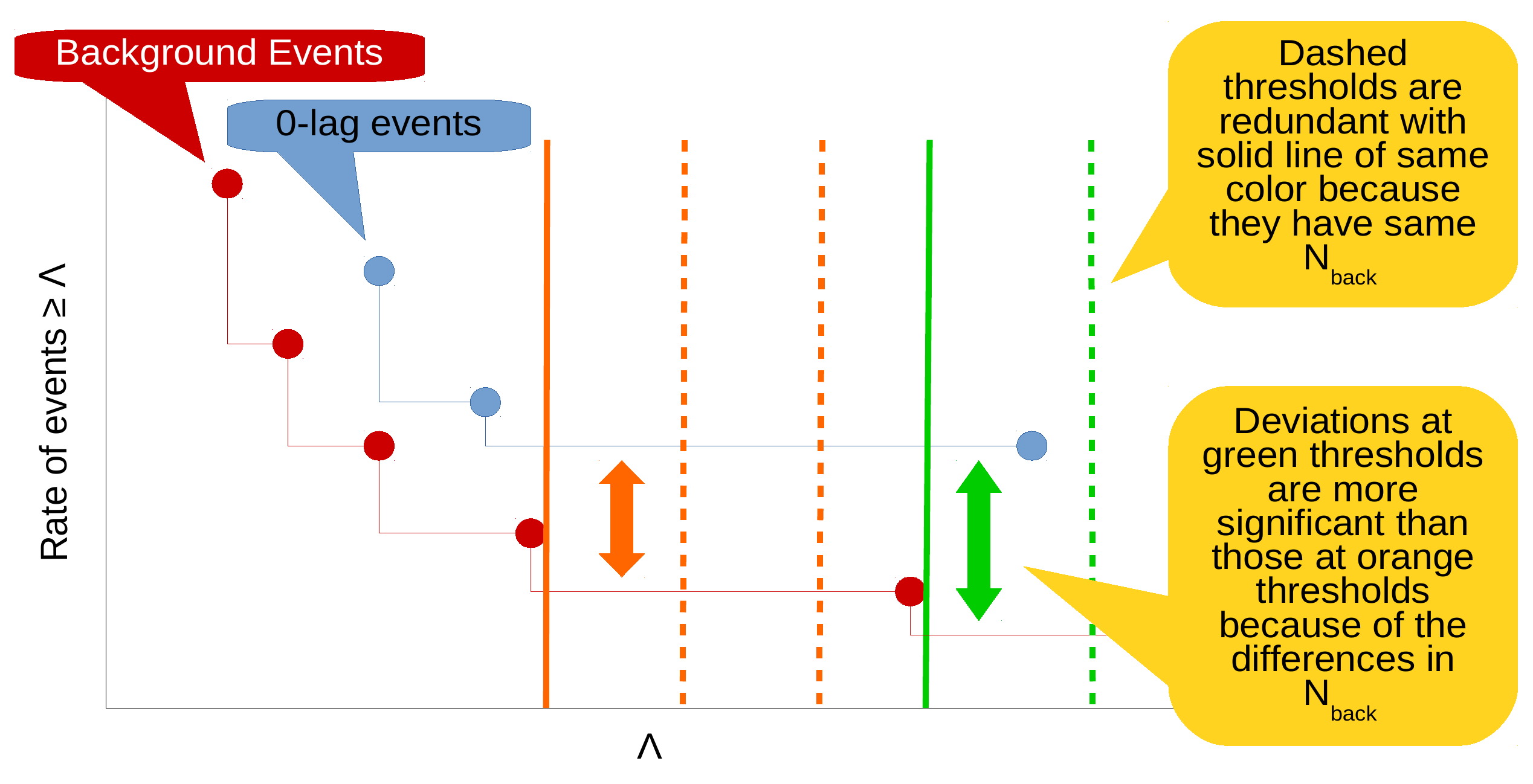}}
    \caption{A graphical explanation of how we should choose to place thresholds in the EST.
    The plot is a toy model of Fig.~\ref{Fig.CumPlotExample}, where the dots represent the location of events.
    As is described in the figure, the set of all unique thresholds is obtained by placing thresholds just past the background events' $\Lambda$'s. 
    The maximum-significance inconsistency of an $i$-event 0-lag realization is found at the threshold closest to the value of the $i^{th}$ loudest 0-lag event without exceeding it.
    }
    \label{Fig.ChoosingThresholds}
\end{figure}

It is sub-optimal to evaluate the consistency of the 0-lag and background measurements at a set of arbitrarily-defined thresholds $\vec{\Lambda}$, both in terms of detection power and computational efficiency.
We will first show how to optimally choose $\vec{\Lambda}$ in order to maximize the detection power of the EST.
A simple thought experiment shows that the most significant 0-lag excesses occur when a threshold $\Lambda$ takes the exact value of a 0-lag event, as is illustrated in Fig.~\ref{Fig.ChoosingThresholds}.
We will call the occurrence of $i$ 0-lag events being measured to exceed a threshold an ``i-event realization'' of the 0-lag.
We first choose a threshold at a value of $\Lambda$ that has no 0-lag events exceeding it.
We then decrease the value of $\Lambda$ until the threshold reaches a 0-lag event, meaning we now have a single-event realization.
Now let us again decrease $\Lambda$, but only slightly so that there is still only a single 0-lag event exceeding it.
Because of the monotonicity of the cumulative representation, the event rate $\lambda\left(\Lambda\right)$ could only have increased or remain unchanged as we decreased $\Lambda$ from that of the loudest 0-lag event.
As a result, the FAP of the current single-event realization is now greater than the FAP of the single-event realization measured at the loudest 0-lag event itself.
We can continue to decrease $\Lambda$ until we reach the second-loudest 0-lag event.
Following the above argument, we find that the most significant two-event realization occurs at the second loudest 0-lag event.
By induction, we see that the most significant $i$-event realization will occur when the threshold $\Lambda$ is located at the $i^{th}$ loudest 0-lag event.

In addition, we note that there are a finite number of thresholds that give us unique p-values, which importantly means that there is a maximum dimension of $\vec{\Lambda}$ above which no new information can be gained from the measured data.
To prove this, let us revisit the above thought experiment illustrated in Fig.~\ref{Fig.ChoosingThresholds}.
We again place a single threshold $\Lambda$ at the value of the $i^{th}$-loudest 0-lag event.
Let us also assume we do not know the event rate $\lambda\left(\Lambda\right)$ exactly, but instead our estimates of it rely on our background measurements.
The significance estimate at the threshold in this scenario takes the form of Eq.~\ref{Eq.FAP_single_ML},~\ref{Eq.FAP_single_uniform}, or~\ref{Eq.FAP_single_Jeffreys}, depending on the choice of event rate prior.
The relevant background measurement is $N_\text{back}$: the number of measured background events exceeding the threshold $\Lambda$.
We can decrease the value of $\Lambda$ from the $i^{th}$-loudest 0-lag event without changing the significance of the $i$-event realization as long as $N_\text{back}$ remain unchanged.
Thus, if we define $\vec{\Lambda}$ to consist of thresholds located at $\Lambda_n + \epsilon$, where $\Lambda_n$ is the measured value of the $n^{th}$ background event and $\epsilon \rightarrow 0$, for every background event in the measurement, we can calculate the unique significance of every possible 0-lag measurement.

%===
\subsubsection{Calculating the EST p-value}

Now that we have an optimal procedure for choosing unique thresholds in the measurement's tail, we define the EST.
The $k$-threshold EST evaluates the statistical consistency of 0-lag and background measurements at the $k$-sharpest binnings of the 0-lag's high-$\Lambda$ tail.
The use of sharp tail bins ensures that any statistical inconsistencies will involve the $k$-loudest 0-lag events.
The final significance of the EST is given by the p-value
\begin{equation}\label{Eq.EST_ETF}
  \text{FAP}_\text{EST} = \text{ETF} \cdot \text{FAP}_\text{min}
\end{equation}
where $\text{FAP}_\text{min}$ is the minimum FAP observed across all of the $k$ thresholds.
Calculating the p-value of the EST takes the following algorithmic form:
  \begin{enumerate}
  \item  Choose $k$, where we will evaluate the single-threshold FAP at each of the $k$-loudest 0-lag events.
  \item  Evaluate the single-threshold FAP at each of the k-loudest 0-lag events using Eq.~\ref{Eq.FAP_single_ML},~\ref{Eq.FAP_single_uniform}, or~\ref{Eq.FAP_single_Jeffreys}, depending on the choice of rate prior.
  Thus, at the $i^{th}$ threshold, $N_{\text{back}}$ is equal to the number of background events exceeding the $i^{th}$ loudest 0-lag event.
  We also choose $N_\text{0lag}^* = i$ so that we are evaluating the significance of seeing the measured $i$-event 0-lag realization.
  Because we are evaluating significances at the measured 0-lag events, any inconsistencies between the 0-lag and background measurements will have maximum significance.
  \item  We are looking for the most significant 0-lag inconsistencies, so we choose our single-threshold reference FAP to be the minimum single-threshold FAP found across the $k$ thresholds, FAP$_\text{min}$; i.e., we choose $\text{FAP}_\text{single} = \text{FAP}_\text{min}$.
  \item  We finally need to find the ETF that maps our minimum single-threshold FAP into a calibrated multiple-threshold FAP via Eq.~\ref{Eq.EST_ETF}.
  To do this, we need to find the $k$ ``critical'' thresholds $\vec{\Lambda} = \lbrace \Lambda_i : i \in 1,\ldots,k \rbrace$ that produce 0-lag realizations with FAPs closest to, without exceeding, FAP$_\text{min}$ (see Eq.~\ref{Eq.CriticalThresholds}).
  We already saw that \textit{any threshold} is equivalent to some threshold located at a background event.
  Thus, we only need to consider thresholds located at the background events. 
  This feature makes the determination of the critical thresholds independent of the shape of the underlying noise event rate $\lambda\left(\Lambda\right)$.
  We find the $k$ critical thresholds using the following algorithm:
    \begin{enumerate}
    \item  Initialize $n = 0$, $i = 1$, and all critical thresholds as undefined.
    \item  Until $i > k$:
    \begin{enumerate}
      \item\label{Item.RepeatPoint}  Calculate FAP$_\text{test}$ at $N_\text{back} = n$ and $N_\text{0lag}^* = i$ using Eq.~\ref{Eq.FAP_single_ML},~\ref{Eq.FAP_single_uniform}, or~\ref{Eq.FAP_single_Jeffreys}, depending on the choice of prior.
      \item  If $\text{FAP}_\text{test} \leq \text{FAP}_\text{min}$, we have achieved a FAP less than the reference FAP.
      Store the current estimate of the $i^{th}$ critical threshold $\Lambda_i$ to be the value of the $n^{th}$ loudest background event.
      Larger values of $n$ can only produce less-significant $i$-event realizations with larger FAPs, so iterate $n \rightarrow n + 1$.
      \item  If instead $\text{FAP}_\text{test} > \text{FAP}_\text{min}$, we no longer have achieved a FAP less than the reference FAP, meaning we have already passed the true critical threshold $\Lambda_i$.
      Thus, we iterate onto the next critical threshold with $i \rightarrow i + 1$.
      \end{enumerate}
    \end{enumerate}
  We note that it is possible that some critical thresholds remain undefined if there exists no $i$-event 0-lag realization with a FAP less than FAP$_\text{min}$.
  In these instances, we remove the undefined thresholds from $\vec{\Lambda}$ (meaning there might be fewer than $k$ critical thresholds).
  \item  The set of critical values $\vec{\Lambda}$ uniquely defines all possible 0-lag realizations involving the $k$-loudest 0-lag events with FAPs less than or equal to FAP$_\text{min}$.
  Thus, we finally evaluate Eq.~\ref{Eq.FAP_multiple_message} at the critical thresholds $\vec{\Lambda}$ with $N_\text{0lag,i}^* = i$.
  The result is the properly-calibrated p-value of our $k$-threshold EST test:  FAP$_\text{EST}$. 
  \end{enumerate}

%===
\subsubsection{Salient features of the EST}
  
We briefly note two salient features of the EST that give it exceptional power to detect statistical inconsistencies in certain scenarios.
First, it incorporates knowledge of the measurement durations $T_\text{0lag}$ and $T_\text{back}$ into the statistical consistency test.
Thus, it is able to distinguish 0-lag and background distributions with different underlying noise event rates $\lambda\left(\Lambda\right)$ even if they have the same distributional shape as a function of $\Lambda$.
Second, while the necessary specification of $k$ means that the EST is parameterized, this parameterization is intentional since it gives the user control over how much of the 0-lag's high-$\Lambda$ tail to test for consistency.
This feature can prove useful when deviations of the 0-lag from the background are expected to be detectable in high-$\Lambda$ regimes but un-detectable in low-$\Lambda$ regimes.

Together, these features make the EST a powerful test for detecting a non-zero rate of GW events superimposed onto detector data in the form of Eq.~\ref{Eq.RateSuperposition} where $\lambda_\text{noise}\left(\Lambda\right) \gg \lambda_\text{GW}\left(\Lambda\right)$ for small values of $\Lambda$.
This power is due to the ability of the EST to test only the high-$\Lambda$ regime of measurements and to detect the elevated event rate of the 0-lag as compared to the background.
  
%===================================================================================================
\section{Comparison with other statistical tests}\label{Sec.Comparison}  

Here we compare the performance of the proposed EST with several other common statistical consistency tests.
These tests are all null tests in the sense that they are testing the hypothesis that all measured data is produced by noise alone.
The first of these tests is the Loudest Event Test (LET).
The LET only computes the Poisson FAP of the loudest 0-lag event~\cite{LoudestEventRate}; \textit{i.e., it is equivalent to the EST with $k = 1$}.
By construction, the loudest 0-lag event will produce the most significant single-event FAP since the noise event rate $\lambda_\text{noise}\left(\Lambda\right)$ is lowest at this event by definition of the cumulative representation.
The LET is commonly used in GW searches to make detection statements on individual 0-lag events (e.g., see ~\cite{O1BBH}).
In practice, if the FAP of the loudest 0-lag event is found to meet some (strict) significance threshold, the 0-lag event is classified as a GW detection.
It can then be excised from the analysis data set, and the consistency of the remaining 0-lag events with the background measurement can be tested using the new loudest event.
This combination of the LET and GW-detection excision can be repeated until the 0-lag measurement is found to be statistically consistent with the background measurement.
Such an application of the LET is equivalent to evaluating the significance of each 0-lag event in isolation of other events.
This interpretation provides the LET with an important feature:  any statistical inconsistencies found by it are attributed entirely to single events, meaning LET detection statements consist of GW events alone. 
As a comparison, the EST detection statement points to a statistical inconsistency somewhere in the $k$-loudest 0-lag events, but it does not uniquely specify which of these events are GW events and which are noise events.
We have verified that the $k=1$ EST produces results that are identical to those of the LET.

In the opposite extreme, we can also consider consistency tests that compare the empirical distributions of two measurements in their entirety.
Two common distributional tests are the Kolmogorov-Smirnov (KS) Test~\cite{Kolmogorov,Smirnoff} and the Anderson-Darling (AD) Test~\cite{AndersonDarling52,AndersonDarling54,KSampSDTest}.
The KS test is most sensitive to inconsistencies in the bulks of the empirical distributions, while the AD test is more sensitive to inconsistencies in the tails~\cite{KSvsAD}.
Both of these test are nonparametric in the sense that they always test the entire distribution in the same way and cannot be tuned to focus on specific regions of the distribution.
As a result, any detection statements made with the KS or AD tests provide no information as to which 0-lag events are GW events and which are noise events.
We also note that both the KS test and the AD test only compare the normalized distributional shapes of the 0-lag and background measurements, i.e., they do not incorporate the relative event rates of the 0-lag and background measurements.

We see that the EST strikes a middle ground between the tradeoffs of the LET and the KS/AD tests.
The LET incorporates the relative 0-lag and background event rates and pinpoints statistical inconsistencies to single events, but it only tests consistency at a single 0-lag event.
The KS and AD tests evaluate the consistency between the entire empirical distributions of the 0-lag and background measurements, but they do not incorporate relative event rate information and cannot pinpoint which 0-lag events are the GW events causing the distributional inconsistencies.
The EST incorporates relative event rate information, tests consistency at the user-specified $k$-loudest 0-lag events, and confines any statistical inconsistencies to these same $k$ 0-lag events.
In the remainder of this section, we will compare these different tests and highlight the effects of the above tradeoffs. 
The first comparison is the calibration of these tests, i.e., whether the p-values reported by each test occur at the expected frequencies.
The second comparison is the detection efficiency of the tests, i.e., how often 0-lag measurements containing GW events are deemed inconsistent with the null hypothesis.

In order to perform these comparisons, we draw 10 random background measurements from the noise event distributions estimated by (1) the burst pipeline oLIB~\cite{oLIBMethods} and (2) the CBC pipeline PyCBC~\cite{PyCBCAlgo,PyCBCCode} in O1~\cite{O1AllSky,O1BBH}.  
For each of these 10 background measurements, we sample $n$ noise events from the estimated noise distribution, where $n$ itself is drawn from a Poisson distribution assuming a measurement duration of 1,000 years and a mean event rate of 100 events per year.
For each of these 10 background measurements, we then simulate 10,000 0-lag measurements.
To do so, we first sample $m$ of the noise events from the estimated noise distribution, where $m$ is drawn from a Poisson distribution assuming a measurement duration of 1 year and a mean event rate of 100 events per year.
GW events are then superimposed onto these 0-lag realizations by similarly drawing events from a specified GW event distribution at a given rate and appending them to the list of 0-lag noise events.
Such a methodology allows us to properly characterize the null test 10 different times under the assumption that all events in both the 0-lag and background measurements are Poissonian realizations of the same underlying noise distribution.

%========================
\subsection{Calibration}\label{SubSec.Calibration}

\begin{figure}[t!]
    \textbf{GW Bursts}\par\medskip
    \scalebox{.7}{\includegraphics[width=0.5\textwidth]{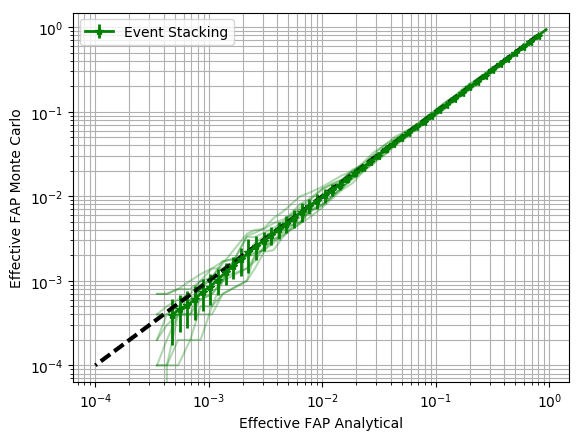}}\par\medskip
    \textbf{CBC}\par\medskip
    \scalebox{.7}{\includegraphics[width=0.5\textwidth]{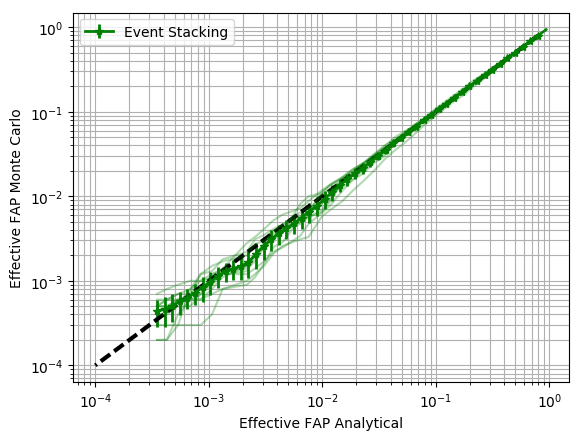}}
    \caption{The calibration plots for the 5-threshold EST with a Jeffreys rate prior for noise-only simulations of a GW burst search (top) and a CBC search (bottom).
    The effective FAP found analytically is the p-value reported by the EST, while the effective FAP found via Monte Carlo is the fraction of the 10,000 0-lag trials that had a p-value less than or equal to the value of the abscissa axis.
    We plot the results for 10 different background measurements (thin continuous lines), along with the mean and standard deviation of these 10 measurements (error bars).
    We find that the EST p-values lie along the expected diagonal (dashed line), meaning it is well-calibrated.
    These results hold for ESTs regardless of the number of thresholds used or choice of rate prior.
    }
    \label{Fig.StackedPPplots}
\end{figure}

\begin{figure}[t!]
    \textbf{Loudest Event Test}\par\medskip
    \scalebox{.7}{\includegraphics[width=0.5\textwidth]{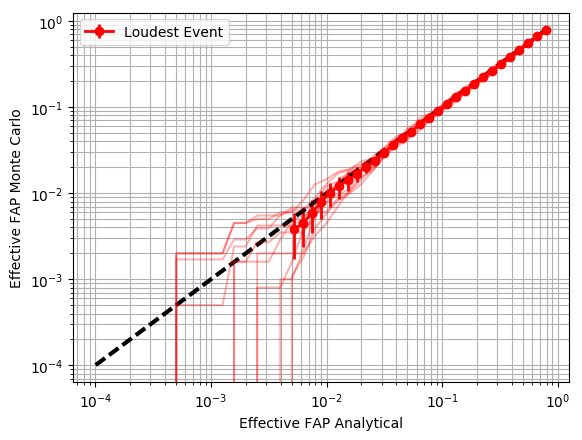}}\par\medskip
    \textbf{KS Test}\par\medskip
    \scalebox{.7}{\includegraphics[width=0.5\textwidth]{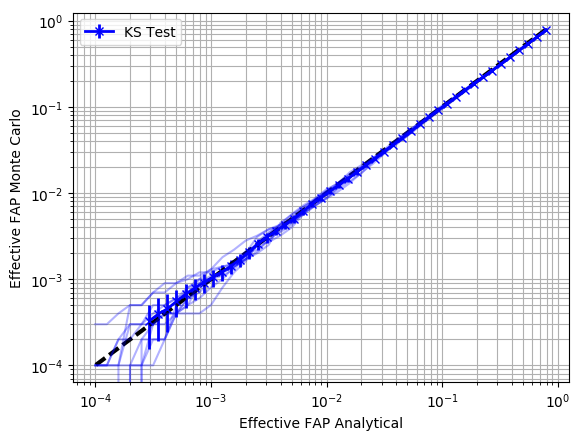}}\par\medskip
    \textbf{AD Test}\par\medskip
    \scalebox{.7}{\includegraphics[width=0.5\textwidth]{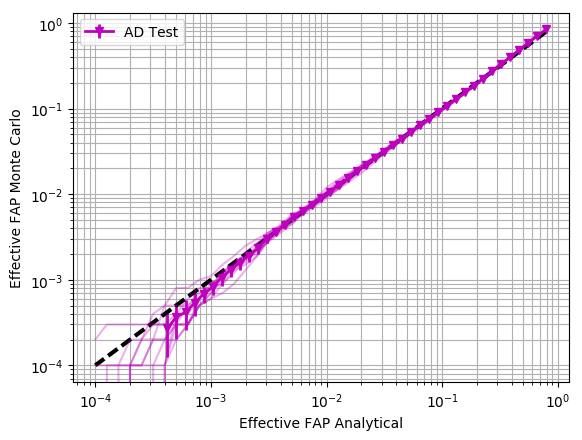}}
    \caption{The calibration plots for the LET with a Jeffreys rate prior (top), KS test (middle), and AD test (bottom) for noise-only simulations of a GW burst search.
    The effective FAP found analytically is the p-value reported by the tests, while the effective FAP found via Monte Carlo is the fraction of the 10,000 0-lag trials that had a p-value less than or equal to the value of the abscissa axis.
    We plot the results for 10 different background measurements (thin continuous lines), along with the mean and standard deviation of these 10 measurements (error bars).
    We find that the p-values of all tests lie along the expected diagonal (dashed line), meaning they are all well-calibrated.
    These results hold for CBC searches as well.
    }
    \label{Fig.OtherPPplots}
\end{figure}

In order to test the calibration of the statistical consistency tests, we run the above simulations without injecting any GW events into the 0-lag, i.e., with $\lambda_\text{GW}\left(\Lambda\right) = 0$ for all $\Lambda$.
Thus, the 0-lag and background measurements should be found to be statistically consistent with respect to the reported p-values of the null test.
In other words, the p-values reported by the null test should be equivalent to the frequency at which noise-only inconsistencies of the same or greater significance occur.

For this work, we will restrict our discussion to the scenario where $T_\text{back} \gg T_\text{0lag}$ since most GW searches operate in this regime.
Analyzing these $\lambda_\text{GW}\left(\Lambda\right) = 0$ simulations, we find that the EST, LET, KS test, and AD test are all well-calibrated in the sense that each p-value statement is reported in that fraction of the 10,000 0-lag trials.
The evidence of this calibration is shown in Figs.~\ref{Fig.StackedPPplots} and~\ref{Fig.OtherPPplots}, where points lie on the predicted diagonal line within the sampling error bars.
We note that in the limit where $\frac{T_\text{back}}{T_\text{0lag}} \lesssim 10$, the mean realization still appears to be well-calibrated, but with large error bars since the background measurement poorly estimates the underlying event rate distribution in this regime.
For the EST, this statement holds invariant of the number of thresholds, the shape of the background distribution, or the choice of rate prior used (i.e., Maximum Likelihood, Uniform, or Jeffreys).
The choice of rate prior does influence the size of the calibration error bars when $T_\text{back} \sim T_\text{0lag}$ and there is a lack of background data to estimate the underlying rate distribution.
However, the effect of the prior choice becomes negligible in the limit $T_\text{back} \gg T_\text{0lag}$ where the same rate estimates become data-dominated.
Because we will only consider $T_\text{back} \gg T_\text{0lag}$ in the remainder of this work, we will only report the results obtained when using the Jeffreys prior.

\begin{figure}[t!]
    \textbf{k=10 Thresholds}\par\medskip
    \scalebox{.7}{\includegraphics[width=0.5\textwidth]{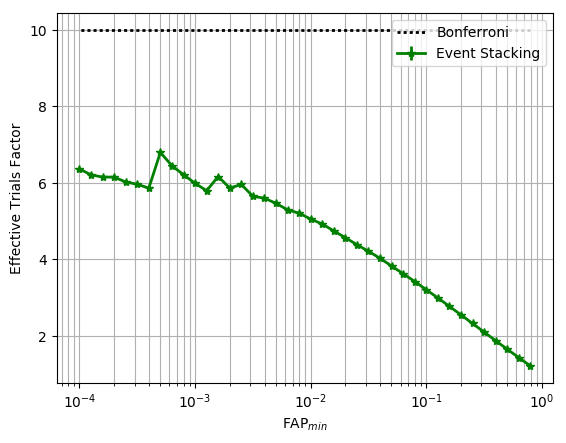}}\par\medskip
    \textbf{k=5 Thresholds}\par\medskip
    \scalebox{.7}{\includegraphics[width=0.5\textwidth]{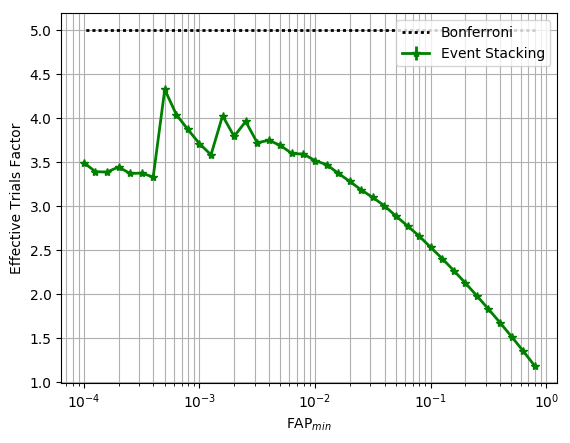}}\par\medskip
    \textbf{k=3 Thresholds}\par\medskip
    \scalebox{.7}{\includegraphics[width=0.5\textwidth]{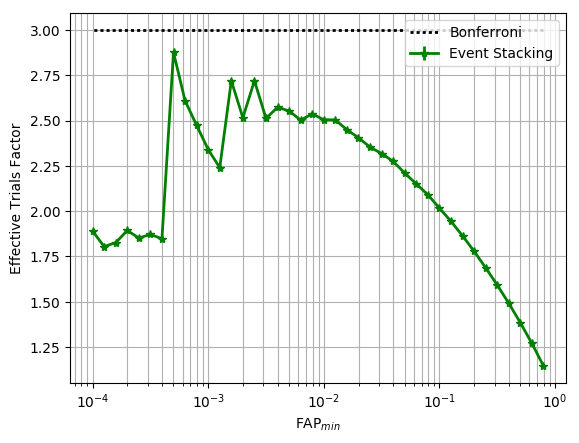}}
    \caption{The effective trials factor (ETF) as a function of the minimum single-threshold FAP observed (FAP$_\text{min}$) for the ESTs using 10 thresholds (top), 5 thresholds (middle), and 3 thresholds (bottom).
    We also plot the naive trials factor of $k$ that would be obtained by applying the Bonferroni correction to a $k$-threshold test.
    We note that the effective trials factor can be significantly lower than the Bonferroni correction, especially for ESTs using a large number of thresholds.
    The jagged nature of the curve at low FAPs is due to the discrete number of background events used to estimate the underlying rate distribution.
    The steep dropoff of $\sim 1$ that occurs around a FAP$_\text{min}$ of $4 \times 10^{-4}$ corresponds to the removal of single-event realization thresholds from the EST.
    These results hold independent of the underlying event distribution $\lambda_\text{noise}\left(\Lambda\right)$ for given measurement durations $T_\text{0lag}$ and $T_\text{back}$.
    }
    \label{Fig.StackedTFplots}
\end{figure}

We note that the calibration of the EST is only possible due to the effective-trials-factor adjustment derived in Sec.~\ref{SubSec.MultipleThresholds}.
Fig.~\ref{Fig.StackedTFplots} compares the effective trials factor, ETF, used in Eq.~\ref{Eq.EST_ETF} for the k-threshold EST to the naive Bonferroni correction that would have $\text{ETF} = k$ across all FAPs.
Note that the Bonferroni correction over-estimates ETF across all FAPs, and thus it over-estimates the rate at which the noise-only statistical inconsistencies occur.
As a result, it under-estimates the significance of any inconsistencies.
The jagged nature of the ETF curve is a result of two effects.
The first is the discrete nature of the 0-lag and background measurements, which prevents us from selecting a set of $k$ critical thresholds whose respective realization probabilities perfectly match $\text{FAP}_\text{min}$.
As discussed in Sec.~\ref{SubSec.EST}, each threshold's FAP merely approximate $\text{FAP}_\text{min}$ as closely as possible.
These approximations are more accurate, and thus the ETF curve is smoother, at high FAPs than at low FAPs since $N_\text{back}$ is larger at high FAPs and the estimates of the underlying rate distribution $\lambda_\text{noise}\left(\Lambda\right)$ become more fine-grained.
Second, thresholds corresponding to $i$-event realizations can drop out of the ETF calculation for small $i$ and low FAPs, as discussed in Sec.~\ref{SubSec.EST}.
For example, single-event 0-lag realizations cannot have FAPs much lower than $\frac{T_\text{0lag}}{T_\text{back}}$ (obtained when N$_\text{back} = 1$, $T_\text{back} \gg T_\text{0lag}$, and the Maximum Likelihood prior is used), meaning the single-event threshold is undefined for extremely-low values of FAP$_\text{min}$.
Referencing the Bonferroni approximation, this means the ETF decreases by $\sim 1$ every time a $i$-event realization drops out of the ETF calculation.
Such an abrupt drop in the value of ETF can be seen in Fig.~\ref{Fig.StackedTFplots} around a FAP$_\text{min}$ of $4 \times 10^{-4}$.

%========================
\subsection{Detection Efficiency}\label{SubSec.Efficiency}

We now explore how powerful these tests are in terms of detecting GW events.
Since the tests are all well-calibrated, we can trust their reported p-values to be accurate representations of the rate of noise fluctuations.
We will count as a detection any 0-lag realization where the p-value is less than or equal to the desired FAP threshold of the detection statement.
We note that a 0-lag measurement is either counted as a detection or as a non-detection; we do not take into account how many GW events in the 0-lag are correctly identified as detections.
We will consider two distributions for the injected GW events:  (1) a ``point'' source of GWs, and (2) a uniform-in-volume distribution of GW sources.
When comparing the results for GW burst and CBC searches, we emphasize that GW burst searches typically have backgrounds with heavier high-$\Lambda$ tails.
These tails are present because GW burst searches inherently place minimal constraints on their target GW signals, meaning it is much easier for noise transients to achieve large values of $\Lambda$ in GW burst searches than in the heavily-modeled CBC searches.

%===
\subsubsection{Point sources}\label{SubSubSec.PointSources}

We define the point source distribution to be a set of GW events that always occur at roughly the same value of $\Lambda$\footnote{Physically-motivated point sources can exhibit $\mathcal{O}(1)$ variations in $\Lambda$ as a result of changes in the detectors' antenna responses that occur if the relative position of the (same) source and the detectors changes.}.
While this distribution could be used to model the physically-motivated scenarios of GW events coming from a single matter-dense region (such as the galactic center or a nearby galaxy) or a randomly repeating source, it is more broadly representative of any scenario where we expect events to bunch around a value of $\Lambda$.
One realistic example is the residual of a LET search after all high-significance GW events are detected and removed from the 0-lag.
Since the single-threshold LET can only claim detections above a certain critical value of $\Lambda$, we may expect a set of low-significance events to be located slightly below this value of $\Lambda$.

\begin{figure}[t!]
    \fbox{\scalebox{.45}{\includegraphics[width=0.47\textwidth]{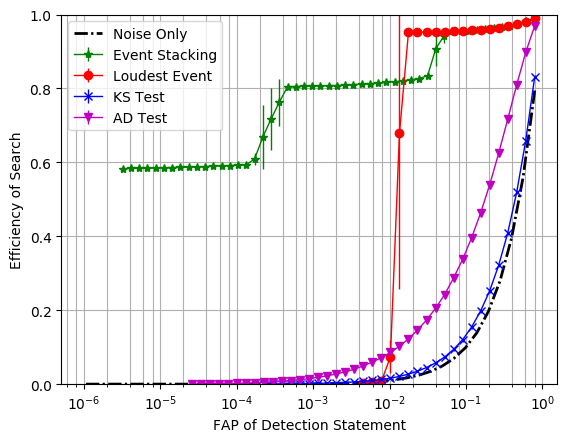}}}
    \fbox{\scalebox{.5}{\includegraphics[width=0.47\textwidth]{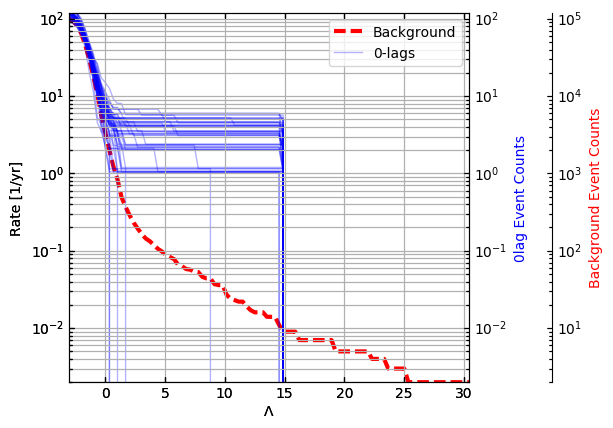}}}
    
    \fbox{\scalebox{.45}{\includegraphics[width=0.47\textwidth]{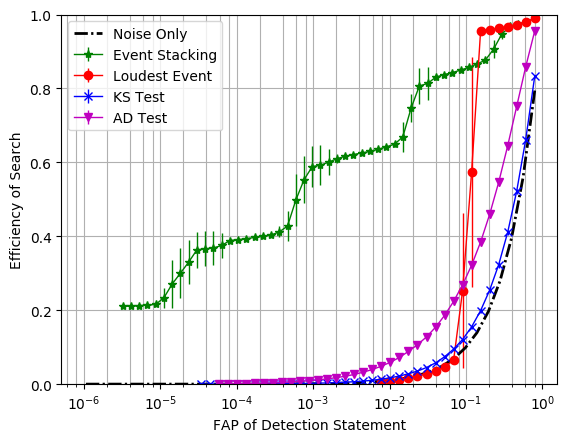}}}
    \fbox{\scalebox{.5}{\includegraphics[width=0.47\textwidth]{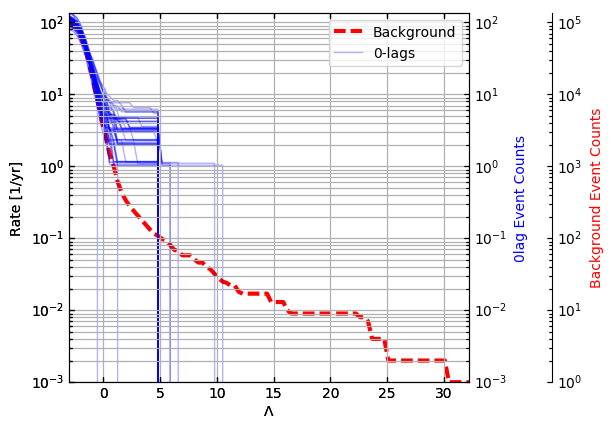}}}
    
    \fbox{\scalebox{.45}{\includegraphics[width=0.47\textwidth]{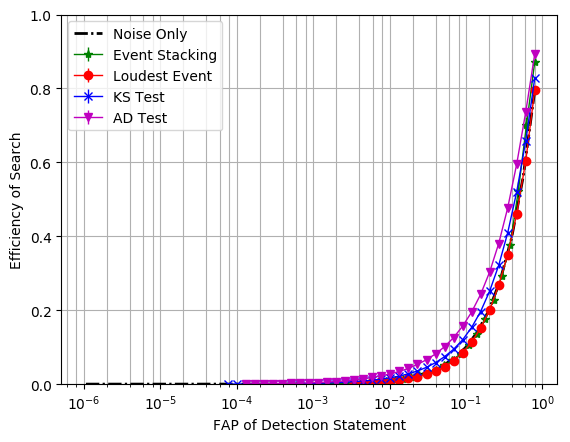}}}
    \fbox{\scalebox{.5}{\includegraphics[width=0.47\textwidth]{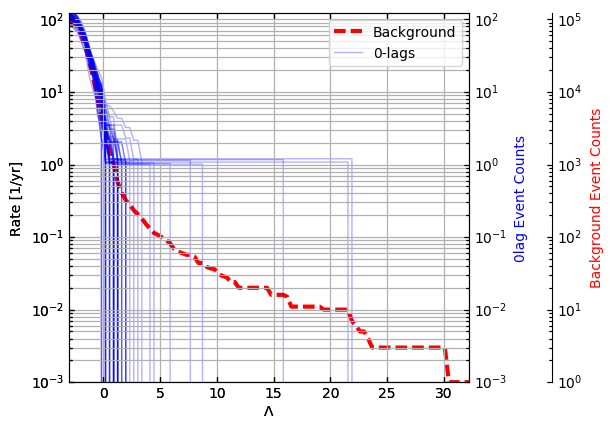}}}
    
    \caption{Left: The ROC curves comparing the GW detection efficiency of the different consistency tests as a function of the effective FAP of their detection statements.
    Here, the GW events are injected into a GW burst search from a ``point'' source distribution.
    The GW events are added to the 0-lag at an average event rate of 3 events per year and have values of $\Lambda =$ 15 (top), 5 (middle), and 0 (bottom).
    The EST uses 5 thresholds, and the $i^{th}$-highest level in the EST curve corresponds to the regime where $i$-event realizations last contribute to detection statements.
    Right: the same plot as Fig.~\ref{Fig.CumPlotExample}, but with the point-source GW events added to the 0-lag measurements.
    }
    \label{Fig.EfficiencyPointSourcesBurst}
\end{figure}

In Fig.~\ref{Fig.EfficiencyPointSourcesBurst}, we explore the detection efficiency of a GW burst search for three different point source distributions that each result in narrowly-varying values of $\Lambda$:  $\Lambda =$ 0, 5, or 15, each at a rate of 3 events per year.
We note that for point sources where $\Lambda = 0$, none of the consistency tests have much detection power, as the number of detections claimed is identical to the number expected from noise fluctuations alone.
When $\Lambda =$ 5 or 10, neither the KS test or AD test have much detection power at any FAP since the distributional inconsistencies only make up a small fraction of the total number of 0-lag events ($\sim 3\%$ on average).
For these scenarios, the LET has good detection power down to a cutoff FAP and has no detection power below this cutoff FAP.
The cutoff FAP corresponds to the FAP of a single-event realization at $\Lambda$.
This behavior helps emphasize the shortcomings of the LET:  the significance of the statistical inconsistency is always defined by the loudest 0-lag event alone, no matter how many GW events may be present in the 0-lag data.
The EST does not have this shortcoming and has substantial detection power across all FAPs for $\Lambda =$ 5 and 10 exactly because it tests for events that are bunched together at similar values of $\Lambda$.
The $i^\text{th}$ highest level seen in the EST detection efficiency curve for the EST represents detection statements being made using $i$-event realizations.
Every time the desired FAP can no longer be reached with $i$-event realizations, the detection efficiency drops to that of the $i+1^{th}$ level since now $i+1$-event realizations are needed to reach the desired FAP.
We note that the $1^{st}$ level of the EST overlaps with the detection efficiency of the LET, which we expect since the LET is equivalent to the $k=1$ EST.
The only difference between the two is that the $1^{st}$ level of the EST ends at slightly higher FAPs than for the LET as a result of the effective-trials-factor penalty applied to the EST (see Eq.~\ref{Eq.EST_ETF}).

\begin{figure}[t!]
    \fbox{\scalebox{.45}{\includegraphics[width=0.47\textwidth]{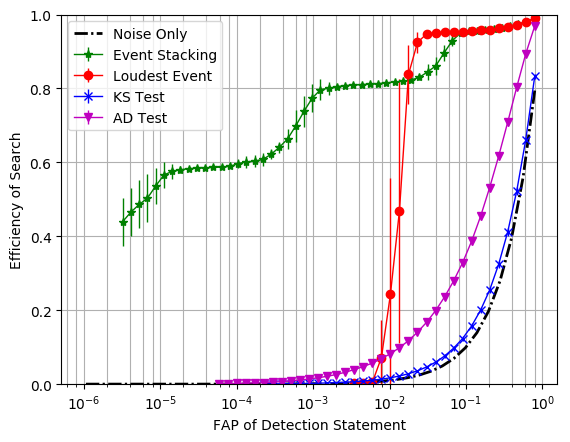}}}
    \fbox{\scalebox{.5}{\includegraphics[width=0.47\textwidth]{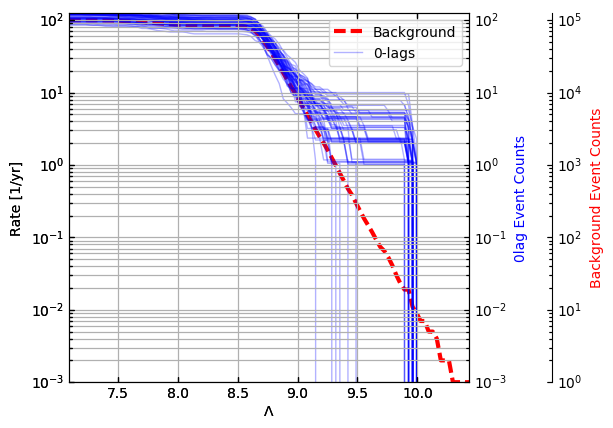}}}
    
    \fbox{\scalebox{.45}{\includegraphics[width=0.47\textwidth]{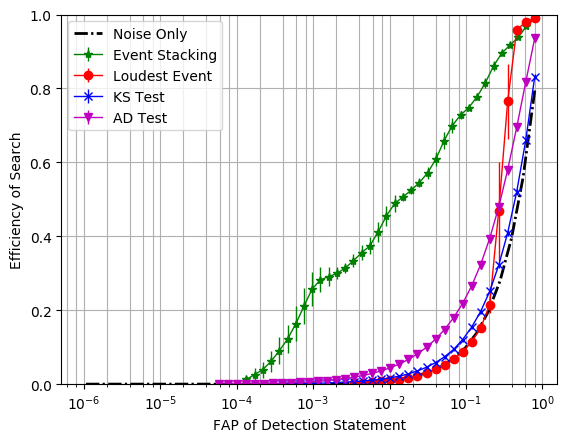}}}
    \fbox{\scalebox{.5}{\includegraphics[width=0.47\textwidth]{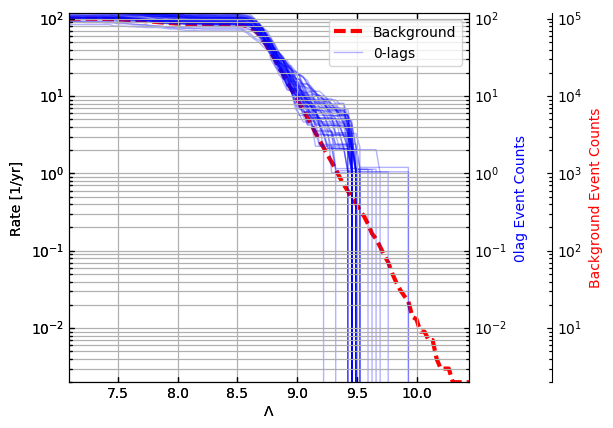}}}
    
    \fbox{\scalebox{.45}{\includegraphics[width=0.47\textwidth]{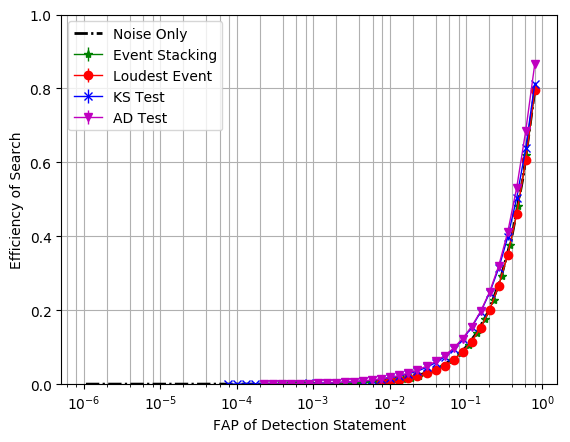}}}
    \fbox{\scalebox{.5}{\includegraphics[width=0.47\textwidth]{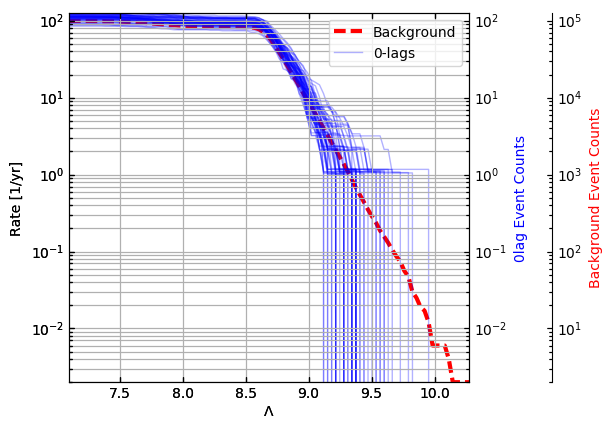}}}
    
    \caption{The same plots as Fig.~\ref{Fig.EfficiencyPointSourcesBurst}, but for a CBC search.
    The GW events are added to the 0-lag at an average event rate of 3 events per year and have values of $\Lambda =$ 10 (top), 9.5 (middle), and 9 (bottom).
    }
    \label{Fig.EfficiencyPointSourcesCBC}
\end{figure}

We note that because all statistical inconsistencies occur at a single value of $\Lambda$ for point sources, the extent of the search background's tail does not affect the behavior of the consistency tests.
All that matters is the relative rate of GW events to noise events at that value of $\Lambda$.
Thus, we observe similar behavior for CBC searches as we do for GW burst searches.
The CBC search results are shown in Fig.~\ref{Fig.EfficiencyPointSourcesCBC} for three different point source distributions:  $\Lambda =$ 9, 9.5, or 10, each at a rate of 3 events per year.
We consider the narrower range of $\Lambda$ values for the CBC search ([9,10]) than for the GW burst search ([0,15]) because the rate of background events falls off much more quickly as a function of $\Lambda$ for the CBC search.  

%===
\subsubsection{Uniform-in-volume sources}\label{SubSubSec.UniformInVolumeSources}

The uniform-in-volume GW source distribution describes the astrophysically-motivated scenario where sources are distributed homogeneously in the spatial volume of the local universe.
If a GW search were to use the optimal matched-filter signal-to-noise ratio (SNR) as its search statistic, namely $\Lambda = \text{SNR}$, we would expect the cumulative GW event distribution to scale $\propto \Lambda^{-3}$ for events located in the low-redshift universe~\cite{Schutz,UniversalSNR,SalvoSchutz}.
It has also been shown~\cite{LynchEMThresholds} that the cumulative noise distribution falls off nearly exponentially as a function of $\Lambda$ in the distribution's bulk. 
These two models imply that the GW event distribution will be more heavily-tailed than the noise distribution.
However, for reasons such as the non-Gaussianity of the LIGO-Virgo detector noise, real GW searches use a combination of SNR and signal-consistency constraints as their search statistics~\cite{AllenChiSq,PyCBCBack,GSTLALLRT,oLIBMethods,cwb2g,BWcomplexity}.
Thus, it is difficult to provide an exact analytical model of the GW event distribution as a function of $\Lambda$ for many real searches.
Instead, we can model the GW event distribution using Monte Carlo simulations where we inject GW signals from a uniform-in-volume source distribution into the LIGO-Virgo data streams.
Running the GW search algorithms over these injections and measuring $\Lambda$ for each, we get an empirical estimate of the GW event distribution as a function of $\Lambda$.
For the CBC search we perform this Monte Carlo procedure using the binary-black-hole waveforms processed by PyCBC for the O1 rates calculation~\cite{O1BBH}, and for the GW burst search we use the sine-Gaussian waveforms used to train the oLIB analysis for the O1 short-duration GW burst search~\cite{O1AllSky}.

\begin{figure}[t!]
    \fbox{\scalebox{.45}{\includegraphics[width=0.47\textwidth]{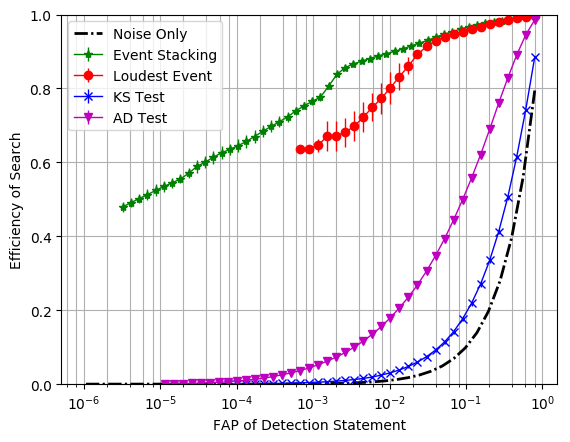}}}
    \fbox{\scalebox{.5}{\includegraphics[width=0.47\textwidth]{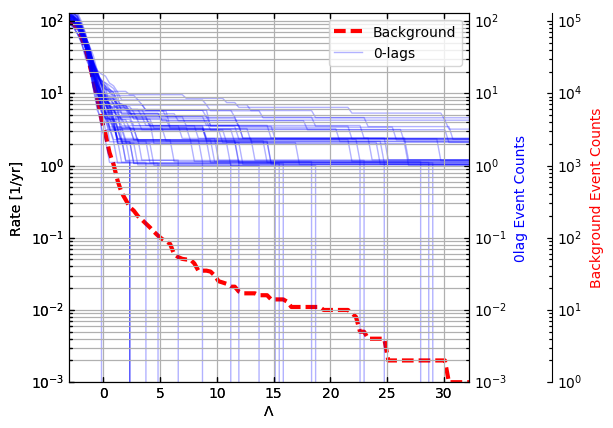}}}
    
    \fbox{\scalebox{.45}{\includegraphics[width=0.47\textwidth]{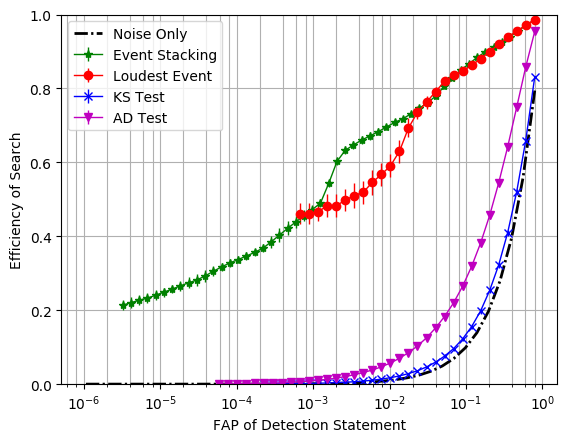}}}
    \fbox{\scalebox{.5}{\includegraphics[width=0.47\textwidth]{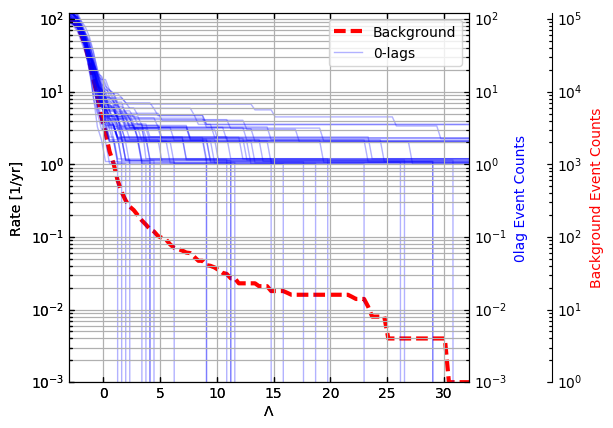}}}
    
    \fbox{\scalebox{.45}{\includegraphics[width=0.47\textwidth]{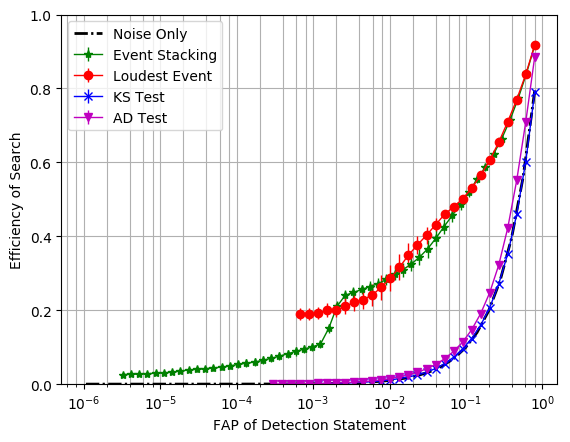}}}
    \fbox{\scalebox{.5}{\includegraphics[width=0.47\textwidth]{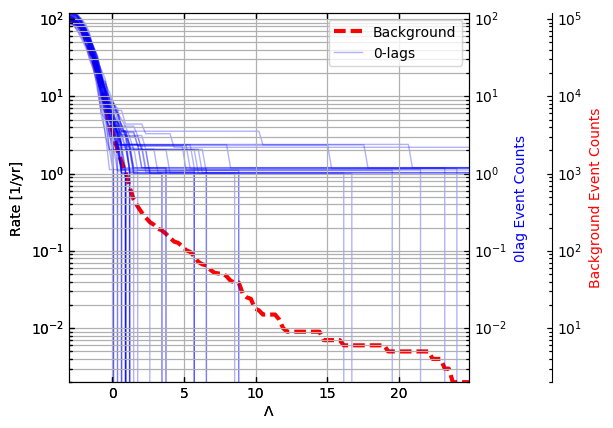}}}
    
    \caption{Left: The ROC curves comparing the GW detection efficiency of the different consistency tests as a function of the effective FAP of their detection statements.
    Here, the GW events are injected into a GW burst search from a uniform-in-volume distribution.
    The GW events are added to the 0-lag at average event rates of 5 events per year (top), 3 events per year (middle), and 1 event per year (bottom).
    The EST uses 5 thresholds, and the $i^{th}$-highest level in the EST curve corresponds to the regime where $i$-event realizations last contribute to detection statements.
    Right: the same plot as Fig.~\ref{Fig.CumPlotExample}, but with the uniform-in-volume GW events added to the 0-lag measurements.
    }
    \label{Fig.EfficiencyUIVSourcesBurst}
\end{figure}

In Fig.~\ref{Fig.EfficiencyUIVSourcesBurst}, we first explore the detection efficiency of a GW burst search for three different uniform-in-volume distributions: GW event rates of 1, 3, and 5 events per year.
The KS and AD tests do not have much detection power at any GW event rate.
This weakness is observed because the noise event rate is 100 events per year, so the statistical inconsistencies only make up $\sim 1 - 5 \%$ of the empirical distributions.
As seen in the example 0-lag realizations of Fig.~\ref{Fig.EfficiencyUIVSourcesBurst}, the differences in relative event rates between the 0-lag and background measurements are most evident in the high-$\Lambda$ tail.
As a result, the EST and LET have much more significant detection power than either the KS or AD tests for all GW event rates, precisely because they test for inconsistent background and 0-lag event rates in the high-$\Lambda$ tail.
At low GW event rates, single-event 0-lag realizations at large values of $\Lambda$ provide the most significant inconsistencies, and thus the LET and EST have similar detection efficiencies.
However, as the GW event rate increases, multiple-event 0-lag realizations are more likely to occur at large values of $\Lambda$, which the EST can detect but the LET cannot.
Thus, EST begins to noticeably outperform the LET as the GW event rate increases.
We also note that because single-event 0-lag realizations can only be detected to FAPs $\sim \frac{T_\text{0lag}}{T_\text{back}}$, the minimum FAP achievable by the LET has a hard-cutoff set by the duration of the background measurement.
Here, the cutoff occurs at a FAP of about $6 \times 10^{-4}$, consistent with our estimates.
On the other hand, the FAP of multiple-event realizations can be much lower than the FAP of single-event realizations.
This feature gives the EST the ability to make detection statements at much lower FAPs than the LET, even when the background measurement duration is limited.

\begin{figure}[t!]
    \fbox{\scalebox{.45}{\includegraphics[width=0.47\textwidth]{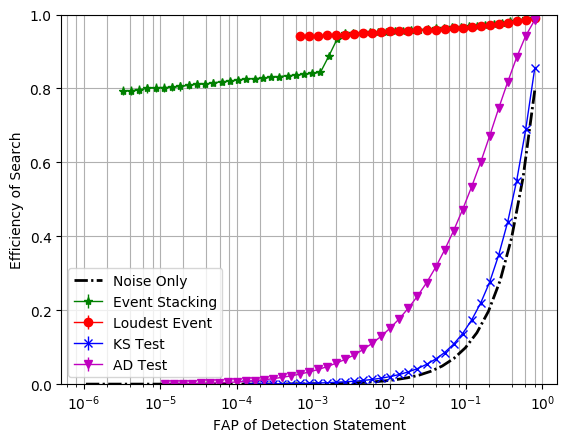}}}
    \fbox{\scalebox{.5}{\includegraphics[width=0.47\textwidth]{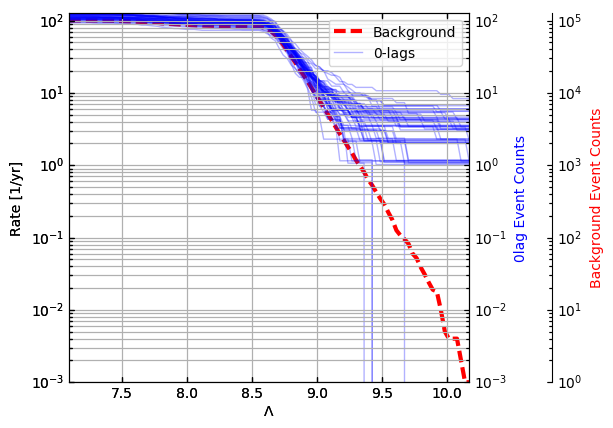}}}
   
    \fbox{\scalebox{.45}{\includegraphics[width=0.47\textwidth]{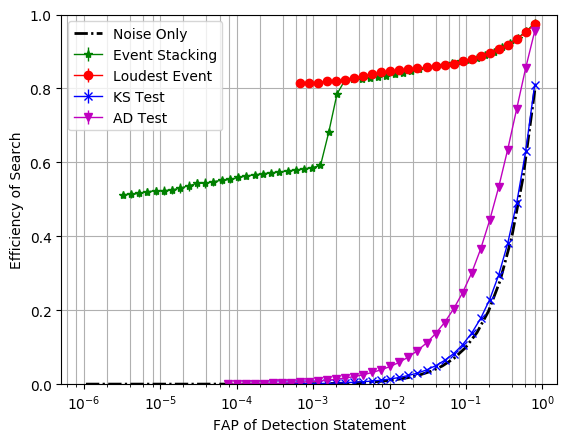}}}
    \fbox{\scalebox{.5}{\includegraphics[width=0.47\textwidth]{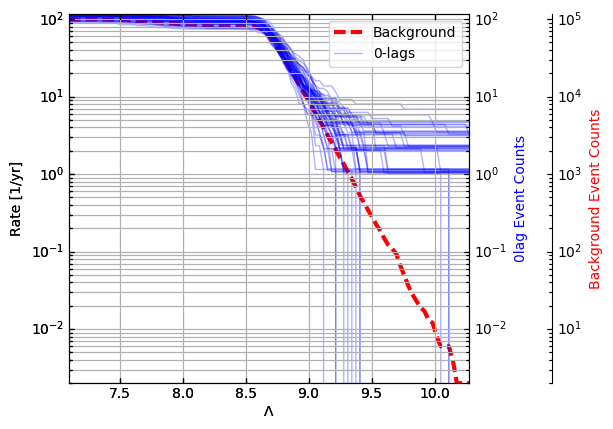}}}
    
    \fbox{\scalebox{.45}{\includegraphics[width=0.47\textwidth]{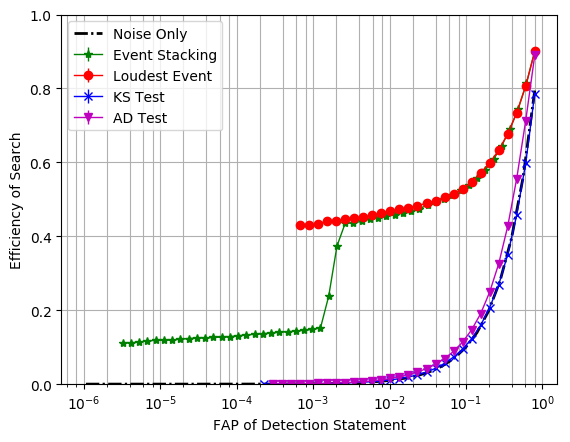}}}
    \fbox{\scalebox{.5}{\includegraphics[width=0.47\textwidth]{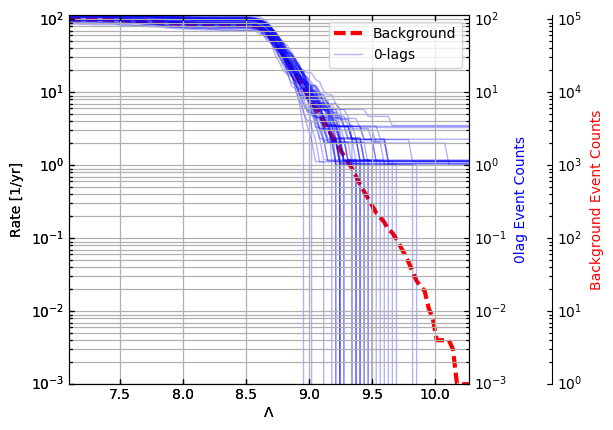}}}
       
    \caption{The same plots as Fig.~\ref{Fig.EfficiencyUIVSourcesBurst}, but for a CBC search.
    The GW events are added to the 0-lag at average event rates of 5 events per year (top), 3 events per year (middle), and 1 event per year (bottom).
    }
    \label{Fig.EfficiencyUIVSourcesCBC}
\end{figure}

We observe similar results, with a few important differences, in Fig.~\ref{Fig.EfficiencyUIVSourcesCBC} for the CBC search, again for three different uniform-in-volume distributions:  GW event rates of 1, 3, and 5 events per year.
The first key difference as compared to the GW burst search is that the detection efficiencies for the LET and EST are very flat as a function of FAP.
These level regions correspond to the regions where $i$-event realizations last contribute to the FAP calculation (e.g., there is only a single level for the LET since it only tests for single-event realizations).
The reason these regions are so flat is because the CBC background is extremely steep as compared to the GW burst background.
Thus, with high probability, the loudest GW event occurs in a region with either (1) very high noise event rates ($\Lambda \lesssim 9$), or (2) very low noise event rates ($\Lambda \gtrsim 10$).
As a result, the loudest GW event is either undetectable or a very significant detection, with little room for ambiguity.
The same holds true for the $i$-loudest 0-lag events.
Because there is little room in the $\Lambda$ parameter space ($\sim$ 1 unit) where the $i$-loudest 0-lag events are detectable together but not in isolation, the EST and LET perform almost identically at FAPs achievable by the LET (i.e., greater than $6 \times 10^{-4}$).
In other words, if an $i$-event realization meets a FAP threshold achievable by the LET, then the loudest-event realization almost certainly meets this FAP threshold as well.
We finally note that the difference in the single-event realization FAP cutoff for the LET and EST is due to the effective-trials-factor penalty applied to the EST (see Eq.~\ref{Eq.EST_ETF}).

%===================================================================================================
\section{Example Application}\label{Sec.UpperLimits}  

Sec.~\ref{Sec.Comparison} demonstrates the relative merits of the EST as compared to the other statistical consistency tests, especially in the case of GW burst searches.
We now explore one physically-motivated application of the EST:  improving the upper limits set by a GW burst search after a null-detection.

%========================
\subsection{Calculating GW Burst Rate-Density Upper Limits}

GW burst searches have not yet detected any non-CBC GW events~\cite{S6Burst,O1AllSky}.
Thus, the results of all GW burst searches to date has been placing rate-density upper limits on unknown (i.e., non-CBC) sources of GW transients.
The classical upper limit on the expected number of GW events in the data, $\mathbb{E} \left[ N_\text{GW} \right]$, placed by an experiment where no inconsistency is observed exceeding the detection significance threshold satisfies
\begin{equation}\label{Eq.DefUpLimit}
  P\left( \text{no significant inconsistency} \ | \ \mathbb{E} \left[ N_\text{GW} \right] \right) \geq 1 - \alpha
\end{equation}
where $\alpha$ is the significance of the confidence interval~\cite{NeymanCL}.
For this example, we will use a FAP detection threshold of $3 \times 10^{-3}$, meaning we can assume that at least one event in any detected inconsistency is a GW event since the noise contamination will be low.
We will also choose to construct the $\alpha = 90\%$ confidence interval.

In order to map the upper limit on $\mathbb{E} \left[ N_\text{GW} \right]$ into a rate-density upper limit, we need to make some assumptions about our astrophysical source distribution.
Assuming that all GWs are emitted from their sources with energy $E_\text{GW}$ at a monochromatic frequency of $f_0$, the distance $d$ of the GW source is given by
\begin{equation}
  d = \sqrt{ \frac{G E_\text{GW}}{\pi^2 c^3 f_0^2 h_\text{rss}^2} }
\end{equation}
where $h_\text{rss}$ is the average root-squared-sum strain amplitude of the GW events as observed at Earth~\cite{SuttonBurstEnergy}.
For uniform-in-volume sources emitting at constant $E_\text{GW}$ and $f_0$, the {\it differential} rate of GW events will be distributed proportionally to $h_\text{rss}^{-4}$ (since $d \propto \frac{1}{h_\text{rss}}$).
We define $\mathcal{R}$ to be the rate-density of the GW events.
The expected number of GW events louder than some minimum strain amplitude $h_\text{rss,min}$ in a GW search of 0-lag duration $T_\text{0lag}$ is given by
\begin{equation}\label{Eq.AvgNumGws}
  \mathbb{E}\left[N_\text{GW}\right] = \mathcal{R} \times \frac{4}{3} \pi \left( \frac{G E_\text{GW}}{\pi^2 c^3 f_0^2 h_\text{rss,min}^2} \right)^{\frac{3}{2}} T_\text{0lag} \ \ .
\end{equation}
The actual number of GW events, $N_\text{GW}$, with strain amplitude louder than $h_\text{rss,min}$ is then a Poisson-distributed random variable with a mean value of $\mathbb{E}\left[N_\text{GW}\right]$ over the measurement duration of $T_\text{0lag}$.
These $N_\text{GW}$ events are distributed in strain amplitude proportionally to $h_\text{rss}^{-4}$.

The upper limit can be expressed in terms of the rate-density by solving Eq.~\ref{Eq.DefUpLimit} to find the $\mathcal{R}$ at which there is probability $1-\alpha$ of detecting no significant inconsistency (where $\mathbb{E} \left[ N_\text{GW} \right]$ is a function of $\mathcal{R}$).
Typically, the LET is used to evaluate the significance of events, where detections are iteratively removed from the 0-lag measurement until it is consistent with the background measurement.
This procedure is equivalent to evaluating all 0-lag events in isolation, meaning the number of detections will be Poisson-distributed and the desired $\mathcal{R}$ can be found by explicitly integrating over the detection efficiency of the GW search on a single-event basis~\cite{S6Burst}.
However, the detection statements of the EST are not performed on isolated events but rather on collective multi-event 0-lag realizations.
As a result, the detection statement of EST is not Poissonian and an analytic calculation of the desired $\mathcal{R}$ is difficult to derive.
Instead, we can use a stochastic approximation method known as the Robbins-Monro algorithm~\cite{RobbinsMonro} to calculate the upper-limit value of $\mathcal{R}$.
To perform the Robbins-Monro algorithm we need to define a random variable $F\left(\mathcal{R}\right)$ whose expected value is equal to the left side of Eq.~\ref{Eq.DefUpLimit}.
I.e., we need
\begin{equation}
\begin{split}
  \mathbb{E} & \left[F\left(\mathcal{R}\right)\right] \\
  &= P\left( \text{no significant inconsistency} \ | \ \mathbb{E} \left[ N_\text{GW}\left(\mathcal{R}\right) \right] \right) \ \ .
\end{split}
\end{equation}
This expectation is achieved if we define
\begin{equation}
\begin{split}
  F & \left(\mathcal{R}\right) \\
  &= \bbbone\left( \text{no significant inconsistency} \ | \ \mathbb{E} \left[ N_\text{GW} \left(\mathcal{R}\right)\right]\right) \ \ .
\end{split}
\end{equation}
In order to satisfy Eq.~\ref{Eq.DefUpLimit}, we wish to find the value of $\mathcal{R}$ where $\mathbb{E}\left[F\left(\mathcal{R}\right)\right] = 1 -\alpha$.
Since the probability of detecting no significant inconsistencies monotonically decreases as a function of $\mathcal{R}$, the iterative equation
\begin{equation}\label{Eq.RobbinsMonro}
  \mathcal{R}_{n+1} = \mathcal{R}_{n} + a_n \left[ F\left(\mathcal{R}_{n}\right) - \left(1-\alpha\right)\right]
\end{equation}
has the right form for convergence to our desired rate.
In fact, Robbins and Monro proved that the above iterative equation will converge (as $n \rightarrow \infty$) to the $\mathcal{R}$ at which $\mathbb{E}\left[F\left(\mathcal{R}\right)\right] = 1 -\alpha$ if $a_n \propto \frac{1}{n}$~\cite{RobbinsMonro}.
As a result, we can find the $\alpha$-confidence upper-limit value of $\mathcal{R}$ {\it for any statistical consistency test} by running the following algorithm:
\begin{enumerate}
  \item  Initialize $\mathcal{R}_1$ to some arbitrary starting value. Initializing $\mathcal{R}_1$ closer to the converged value $\mathcal{R}_\infty$ will speed up the convergence.
  \item  Until $\mathcal{R}_n$ is sufficiently converged:
  \begin{enumerate}
    \item  Draw a value of $N_\text{GW}\left(\mathcal{R}_n\right)$ from a Poisson distribution with mean rate $\mathbb{E}\left[N_\text{GW}\left(\mathcal{R}_n\right)\right]$ (see Eq.~\ref{Eq.AvgNumGws}) and measurement duration $T_\text{0lag}$.
    \item  Draw $N_\text{GW}\left(\mathcal{R}_n\right)$ GW events with strain amplitude greater than $h_\text{rss,min}$ from a strain amplitude distribution proportional to $h_\text{rss}^{-4}$.  
    The values of $\Lambda$ for these GW injections should have previously been measured via the GW search.
    \item  Simulate a 0-lag measurement by sampling $N_\text{noise}$ noise events from the background measurement, where $N_\text{noise}$ is drawn from a Poisson distribution with the same mean event rate as the background and a measurement duration of $T_\text{0lag}$. 
    Append the GW events to this simulated 0-lag measurement.
    \item  Evaluate the statistical consistency test (e.g., EST, LET) using the background and 0-lag measurements.  
    Set $F\left(\mathcal{R}_n\right) = 1$ if there is no inconsistency meeting the FAP detection threshold and $F\left(\mathcal{R}_n\right) = 0$ if a significant inconsistency is detected.
    \item  Find $\mathcal{R}_{n+1}$ using Eq.~\ref{Eq.RobbinsMonro} with $a_n = \frac{\mathcal{R}_1}{n}$ and then iterate $n \rightarrow n + 1$.
  \end{enumerate}
\end{enumerate}

%========================
\subsection{Results}

\begin{table*}[t!]
    \centering
    \scriptsize
    \caption{The $90\%$-confidence rate-density upper limits (in Gpc$^{-3}$ yr$^{-1}$) placed on GW burst sources emitting at $E_\text{GW} = 1 M_\odot c^2$, assuming a uniform-in-volume source distribution.
    The upper limits are calculated for 4 sine-Gaussian waveform morphologies, defined by their central frequency ($f_0$) and quality factor ($Q$), and 2 white-noise burst waveform morphologies, defined by their central $f_0$, bandwidth $\Delta f$, and duration $\tau$.
    These waveforms were used in the LIGO-Virgo O1 analysis~\cite{O1AllSky}.
    We calculate the upper limits using both the LET and EST, and give their ratio.
    We find that the EST sets rate-density upper limits that are stricter than those of the LET by $40\%-240\%$.
    {\it We stress that the configuration of our simulations is ad hoc, meaning these rate-density upper limits are non-astrophysical}.
	    }
    \label{Tab.UpperLimitsComparison}
    \begin{tabular}[c]{c||r|r||r}
        \hline
        SG Morphology & LET [Gpc$^{-3}$ yr$^{-1}$] & EST [Gpc$^{-3}$ yr$^{-1}$] & LET-to-EST ratio \\
        \hline
        $f_0 = 153$, $Q = 9$ & 8.5 & 4.8 & 1.8 \\
        $f_0 = 235$, $Q = 100$ & 48 & 25 & 1.9 \\
        $f_0 = 554$, $Q = 9$ & 1300 & 780 & 1.7 \\
        $f_0 = 849$, $Q = 3$ & 8700 & 6300 & 1.4 \\
        \hline
        WNB Morphology & LET [Gpc$^{-3}$ yr$^{-1}$] & EST [Gpc$^{-3}$ yr$^{-1}$] & LET-to-EST ratio \\
        \hline
        $f_0 = 150$ Hz, $\Delta f = 100$ Hz, $\tau = 0.1$ s & 67 & 26 & 2.6 \\
        $f_0 = 300$ Hz, $\Delta f = 100$ Hz, $\tau = 0.1$ s & 920 & 270 & 3.4 \\
    \end{tabular}
\end{table*}

We now compare how well the LET and EST constrain the GW burst rate-density after a null detection.
We draw noise events and GW events from oLIB's analysis of O1 data.
While the background event distributions are the same as published for the LIGO-Virgo O1 short-duration GW burst analysis~\cite{O1AllSky}, here we arbitrarily choose the event rate to be 100 events per year, the measurement durations to be $T_\text{0lag} = 1$ year and $T_\text{back} = 1000$ years, and the FAP detection threshold to be $3 \times 10^{-3}$.
With these arbitrary choices, {\it the rate-density upper limits calculated in this section are not astrophysical and should not be compared to those published for the O1 analysis}~\cite{O1AllSky}.
However, these parameter choices are reasonably close to those of historical GW burst searches, and thus we expect the general trends of our findings to hold for actual GW burst upper-limit calculations.
We note that this FAP detection threshold is chosen so that a 5-threshold EST is able to make detections using its single-event realization threshold (see Fig.~\ref{Fig.EfficiencyUIVSourcesBurst} for the location of the FAP cutoff for these choices of $T_\text{0lag}$ and $T_\text{back}$).
As a result, our findings should change only negligibly if more background were measured so that $T_\text{back}$ increased. 
We evaluate the upper limits at a confidence of $\alpha = 90\%$ for $E_\text{GW} = 1 M_\odot c^2$ and $h_\text{rss,min}$ well below the detection sensitivity of the O1 oLIB search.
We finally note that because the Robbins-Monro algorithm fixes the value of $\mathbb{E}\left[N_\text{GW}\left(\mathcal{R}_n\right)\right]$, Eq.~\ref{Eq.AvgNumGws} lets us rescale the rate-density upper limits for any choice of GW emission energy using $\mathcal{R} \propto E_\text{GW}^{-\frac{3}{2}}$. 

We compute the GW rate-density upper limits for 4 different sine-Gaussian and 2 different white-noise burst signal morphologies used in the LIGO-Virgo O1 analysis~\cite{O1AllSky}, using both the LET and a 5-threshold EST.
We list the results in Table~\ref{Tab.UpperLimitsComparison}.
The LET results obtained with the Robbins-Monro algorithm match those obtained with the historically-used analytic calculation, giving us confidence that the algorithm converges properly.
We note that the EST consistently provides stricter GW rate-density upper limits than the LET, with the level of improvement ranging from $40\%-240\%$ (see Table~\ref{Tab.UpperLimitsComparison}).
These results imply that although the loudest GW events provide the most useful information for inferring the GW rate-density, quieter events do provide a non-negligible amount of additional information.
This ability to use low-significance events to place stricter limits on GW rate estimates has been seen in the sophisticated models used for CBC population inference~\cite{Farr}.
The extent of the EST's improvement depends on how heavily-tailed the GW event distributions are as a function of $\Lambda$ for each GW signal morphology.
The EST shows greater improvement over the LET for less-heavily-tailed distributions since individual GW events are less likely to be obvious high-$\Lambda$ detections and are instead more likely to bunch together at lower values of $\Lambda$.
We finally note that, in the Monte Carlo calculation, the EST detection statements at the rate-density upper limit contained roughly 3 GW events and 0.1 noise events on average.
As a comparison, the LET detection statements contained 2.3 GW events and 0.003 noise events on average, both matching the analytically-predicted values.
These numbers highlight the minor tradeoff that comes with choosing the EST over the LET:  the EST is capable of detecting more GW events per 0-lag realization than the LET, but it also has a larger noise contamination in its detection statement.

%===================================================================================================
\section{Conclusions}\label{Sec.Conclusions}

The Event Stacking Test (EST) is a null hypothesis test that checks for statistical consistency between a background measurement and an analysis measurement.
This test specifically targets the (``loud'') tail of a distribution, evaluating the consistency of the empirical distributions at $k$ thresholds, where $k$ is defined by the user.
The multiple thresholds have the effect of ``stacking'' multiple analysis events into the tail bin.
As a result, up to $k$ of the loudest analysis events can be detected in any measurement, even if none of them are individually significant enough to be detected in isolation.
We have shown that the EST is well-calibrated, so that the significance of the reported level of inconsistency is representative of noise fluctuations.
For carefully constructed GW searches, this proper calibration ensures that any significant inconsistencies are likely caused by the presence of GW events in the analysis data.

Comparing the GW detection power of the EST to that of other statistical consistency tests, such as single-outlier tests like the Loudest Event Test (LET) or nonparametric distributional tests like the Kolmogorov-Smirnov (KS) test or the Anderson-Darling (AD) test, the benefits of the EST become clear.
The EST is a parameterized test in the sense that the user must choose $k$, meaning tuning is needed to find the value of $k$ that maximizes the detection power of the test.
This freedom to tune is beneficial when searching for GWs since it allows the user to interpolate between single-event tests (LET) and fully distributional tests (KS and AD) depending on the exact shape of the noise and GW event distributions.
The EST takes into account the relative rates of the background and analysis measurements at multiple thresholds, meaning it incorporates more information than just the shapes of the empirical distributions.

As a result of these features, we find that the EST robustly outperforms the KS and AD tests in terms of GW event detection, independent of source type or source distribution.
The noise distributions of CBC searches typically have steeply-falling tails, meaning there is limited room in the tail for low-significance outliers to bunch together and form significant inconsistencies.
As a result, the EST performs comparably to the LET in terms of detecting CBC sources in the significance regime where single-event detections can be made (and obviously outperforms the LET in regimes where only multiple-event detections are possible).
On the other hand, GW burst searches typically have more heavily-tailed noise distributions than CBC searches, meaning there is more room in the tail for low-significance events to bunch together and form significant inconsistencies.
Thus, the EST can achieve noticeably higher GW burst detection efficiencies than the LET, even in the significance regime where single-event detections are common.
As a result, the EST can place upper limits on the astrophysical rate-density of GW burst sources that are $40\%-240\%$ stricter than those set using the LET.

We finally note that one shortcoming of the EST is that its detection statement does not explicitly indicate which of the events causing the statistical inconsistency are GW events and which are noise events.
At the $90\%$ rate-density upper limits, calculated using a false-alarm probability detection threshold of 0.003, we found that roughly $3\%$ of the events found in EST detection statements were noise, as compared to $0.1\%$ of events for the LET.
However, this feature should not discourage the use of the EST since any detection of GW bursts would necessarily launch an intense parameter-estimation follow-up in an attempt to explain the events' astrophysical origins.
Using the EST will allow the GW community to detect GW burst events more efficiently and in greater quantity.
These prospects should only encourage the continued development of the parameter estimation tools that will be needed to maximally extract science from what will surely be groundbreaking detections.

%===================================================================================================
\section{Acknowledgments}

The authors acknowledge the support of the National Science Foundation and the LIGO Laboratory.
LIGO was constructed by the California Institute of Technology and Massachusetts Institute of Technology with funding from the National Science Foundation and operates under cooperative agreement PHY-0757058.
We would also like to thank Hsin-Yu Chen, Kwan Yeung Ng, Yiwen Huang, Robert Eisenstein, Satya Mohapatra, Steve Drasco, Reed Essick, and the LIGO-Virgo Burst working group for useful comments and discussion. 
This is LIGO document number P1800170.

%===================================================================================================

\bibliography{refs}

%===================================================================================================
\end{document}